\documentclass{aa} %[twocolumn]{aastex63}
\usepackage{graphicx}
\usepackage{color}
\usepackage{amsmath}
\usepackage{natbib}
\bibpunct{(}{)}{;}{a}{}{,}
\usepackage{siunitx}
\usepackage{float}
\usepackage[caption = false]{subfig}
\usepackage{gensymb}
\usepackage{hyperref}
\usepackage{fancyhdr}
\usepackage{multirow}
\usepackage{tablefootnote}
\usepackage [english]{babel}
\usepackage [autostyle, english = american]{csquotes}
\MakeOuterQuote{"}

\newcommand{\AN}[1]{#1}

\begin{document}

%###############################################################################
%
%     OPENING
%
%###############################################################################
% \thispagestyle{fancy}
% \fancyhf{}
% \rhead{\footnotesize{A modified version of this article was accepted for publication in The Astrophysical Journal}}

\title{Self-Supervised Learning on MeerKAT Wide-Field Continuum Images}

%gpt suggestions...
 % "Exploring Self-Supervised Learning on MeerKAT's Galactic Cluster Legacy Survey: A Radio Astronomy Perspective"
 %    "Unlocking the Potential of Self-Supervised Learning for Multi-Purpose Radio Astronomy Models: A Case Study with MeerKAT"
 %    "Beyond Galaxy Zoo: Self-Supervised Learning on Wide-Field Radio Astronomy Surveys with SSL Frameworks"
 %    "Revolutionizing Radio Astronomy with Self-Supervised Learning: Insights from the MeerKAT Galactic Cluster Legacy Survey"
 %    "From Wide-Field Radio Images to Multi-Purpose Models: Self-Supervised Learning on the SKA-Mid MeerKAT Observations"

%\accepted{30.10.2020}
%\submitjournal{The Astrophysical Journal}

%A modified version of this paper was accepted for publication in The Astrophysical Journal.

%\submitted{A modified version of this article was accepted for publication in The Astrophysical Journal}
%\fancypagestyle{pprintTitle}{%
%\lfoot{}\cfoot{}\rfoot{}
%\renewcommand{\headrulewidth}{0.0pt}
%}

%\correspondingauthor{E. Lastufka}

% \author[0000-0003-1894-2074]{E. Lastufka}
% \affiliation{University of Geneva, Geneva, Switzerland}
% \affiliation{Observatoire de Genève, Université de Genève, 51 Chemin Pegasi, 1290 Versoix, Switzerland}

\author{E. Lastufka\inst{1}, %0000-0003-1894-2074
        O. Bait\inst{2},
        O. Taran\inst{1},
        M. Drozdova\inst{1},
        V. Kinakh\inst{1}, %0000-0001-5301-9141
        D. Piras\inst{1}, %0000-0002-9836-2661
        M. Audard\inst{1}, %0000-0003-4721-034X
        M.Dessauges-Zavadsky\inst{2},
        T. Holotyak\inst{1},
        D. Schaerer\inst{2},
        S. Voloshynovskiy\inst{1}}%\email{erica.lastufka@unige.ch} %      \and

\institute{Department of Computer Science, University of Geneva, 7 route de Drize, 1227 Carouge, Switzerland \email{erica.lastufka@unige.ch}
\and
Department of Astronomy, University of Geneva, 51 Chemin Pegasi, 1290 Versoix, Switzerland}

\date{Received 13 March 2024; accepted 9 August 2024}

%###############################################################################
%
%     ABSTRACT
%
%###############################################################################

  \abstract
  % context heading (optional), max 300 words
{Self-supervised learning (SSL) applied to natural images has demonstrated a remarkable ability to learn meaningful, low-dimension representations without labels, resulting in models that are adaptable to many different tasks. Until now, applications of SSL to astronomical images have been limited to Galaxy Zoo datasets, which require a significant amount of pre-processing to prepare sparse images centered on a single galaxy. With wide-field survey instruments at the forefront of the Square Kilometer Array (SKA) era, this approach to gathering training data is impractical.}
%aims
{We demonstrate that continuum images from surveys like the MeerKAT Galactic Cluster Legacy Survey (MGCLS) can be successfully used with SSL, without extracting single-galaxy cutouts..}
%methods
{Using the SSL framework DINO, we experiment with various preprocessing steps, augmentations, and architectures to determine the optimal approach for this data. We train both ResNet50 and Vision Transformer (ViT) backbones.}
%results
{Our models match state-of-the-art results (trained on Radio Galaxy Zoo) for  FRI/FRII morphology classification. Furthermore, they predict the number of compact sources via linear regression with much higher accuracy. Open-source foundation models trained on natural images such as DINOv2 also excel at simple FRI/FRII classification; the advantage of domain-specific backbones is much smaller models trained on far less data. Smaller models are more efficient to fine-tune, and doing so results in  similar performance between our models, the state-of-the-art, and open-source models on multi-class morphology classification.}
%conclusions
{Using source-rich crops from wide-field images to train multi-purpose models is an easily scalable approach that significantly reduces data preparation time. For the tasks evaluated in this work, twenty thousand crops is sufficient training data for models that produce results similar to state-of-the-art. In the future, complex tasks like source detection and characterization, together with domain-specific tasks, ought to demonstrate the true advantages of training models with radio astronomy data over natural-image foundation models.}

%\keywords{Radio astronomy, SKA, MeerKAT, self-supervised learning, morphology classification}


\titlerunning{MeerKAT SSL}
\authorrunning{E. Lastufka}

\maketitle

\section{Introduction}

%why foundation model
Radio interferometer arrays are infamous for generating large quantities of data, and observations by the Square Kilometer Array (SKA, \citet{dewdney_square_2009}) are expected to reach the exabyte scale. Even once these signals are correlated, processed into calibrated visibilities and then drastically reduced in size via imaging, observations still require further analysis to bring the data to comprehensible levels. Currently, a majority of that work - finding sources and measuring their properties, classification or flagging as an unknown object - is done by experts in pursuit of a particular scientific goal. There is significant overlap between these basic tasks and the capabilities of foundation models in computer vision. These networks are typically trained on massive datasets like ImageNet \citep{deng_imagenet_2009}, and provide a foundation upon which more task-specific models can be built through fine-tuning. 

%why SSL
Self-supervised learning (SSL) is an integral part of training foundation models, as it does not require any labels describing the data in order to learn. Especially in astrophysics, unlabeled data like images or spectra are more abundant than labeled data, such as cutouts of a single galaxy characterized by carefully calculated morphological or spectral parameters. SSL enables models to learn rich, generalized representations from the data itself; its success is evident from works in the field of computer vision such as SwaV \citep{caron_unsupervised_2020}, SimCLR \citep{chen_simple_2020}, BYOL \citep{grill_bootstrap_2020}, DINO v1 and v2 \citep{caron_emerging_2021, oquab_dinov2_2024}, and MSN \citep{assran_masked_2022}. 

Models like these have long since demonstrated incredible success when trained on natural images (e.g. \citet{pathak_context_2016}); they can be used to detect and classify objects, estimate depth, find similar images, and detect copies, among many other tasks. Because general proprieties of images are encoded into a compressed dimension of latent representation, in which similar data points are close in Euclidian space, % --- known as a latent space --- 
models can be quickly fine-tuned to specific tasks and datasets with relative ease. \AN{The ability to adapt to many different tasks is the most important characteristic of a foundation model, although \citet{Bommasani2021FoundationModels} additionally define foundation models as being trained on data with broad characteristics.} 

\AN{The application of foundation models in an out-of-the-box fashion to radio astronomy data has not been fully explored, with many simply assuming the need for domain-specific models because of the unique nature of radio images.
Indeed, images from radio telescopes are fundamentally different from natural images. They} are always reconstructions that start from an inverse Fourier transform of visibilities sampled in the \textit{uv}-plane to the image plane. Weighting of the visibilities determines the trade-off between sensitivity and resolution. The resulting image product contains a number of radio sources, usually sparse relative to the field of view, some amount of noise, and artefacts from the image reconstruction. There can be a large dynamic range between the radio sources or the sources and the noise. Objects of many different scales can be present; most radio sources might be approximated by point sources, but diffuse emission and instrumental effects alike can occupy large areas. %Emission is continuous rather than discrete; 

% 	- imbalanced dataset
% 	- many objects in FOV
% 	- Not object centric
% Objects of very different scales, properties; continuous instead of discrete

Images from optical astronomy share many of these characteristics, and in recent years SSL was applied in this domain by \citet{stein_self-supervised_2021}. Using 42 million images in three bands of individual optical galaxies from the Dark Energy Spectroscopic Instrument (DESI) Legacy Surveys, the authors trained a residual network (ResNet). They utilized an optical-specific set of data augmentations, such as reddening due to galactic extinction and a varying point spread function, to produce different views of a single image.  Contrastive loss was then employed to minimize the distance in latent space between different views, while maximizing this distance between different images. 
 For evaluating the performance of a SSL network, usually a smaller labeled portion of the dataset is used for some downstream task such as classification. With a completely unlabeled dataset, \citet{stein_self-supervised_2021} focused instead on instance-based retrieval, resulting in the Galaxy Finder tool.
\citet{hayat_self-supervised_2021} used a similar approach for Sloan Digital Sky Survey (SDSS) data, which was labelled thanks to Galaxy Zoo. Their self-supervised approach out-performed state-of-the-art supervised models on the tasks of morphology classification and photometric redshift estimation. 

%current progress radio
SSL was brought to radio astronomy by \citet{slijepcevic_learning_2022} and further refined in \citet{slijepcevic_radio_2024}, using the yet-unreleased Radio Galaxy Zoo (RGZ, Wong et al. 2024 in prep) dataset assembled from the VLA FIRST survey \citep{becker_first_1995}. With the Bootstrap Your Own Latent (BYOL) method based on instance differentiation, they were able to find hybrid radio sources using similarity search, as well as exceed baseline supervised binary classification performance. Unlike in the models from previous works mentioned, which used images from optical astronomy, the contrastive loss in BYOL is calculated without negative pairs; only different views of the same image are relevant. 

%why our approach necessary
These latest developments in both optical and radio astronomy have similarities not only in the networks used, but also in the core characteristics of the images. Training data are relatively small, typically 70-150 pixel cutouts centered on a single galaxy, meaning that the images are often very sparse. \AN{This postage-stamp format is a logical choice considering the goal of morphology classification or similarity search. However, it builds a reliance on data products from the very end of a traditional processing pipeline, which could take days or even months to reach.} 
RGZ was assembled using archival survey data, which has the advantage of no additional waiting time for calibrating and processing data arriving from the interferometer; creating a similar dataset \AN{of individual galaxy cutouts} from modern surveys from SKA precursors whose pipelines are not yet fully finalized would be a time-consuming endeavor. %The approach of identifying sources and making individual cutouts for SKA data will be extremely time-consuming, as we can see already from the amount and wealth of observations being gathered by its precursors.

%why our approach is different
The SKA-Mid precursors MeerKAT and Australian Square Kilometer Array Pathfinder (ASKAP) have been operating since 2018 and 2015, respectively. %The data they produce are much different to training data from Galaxy Zoo or RGZ.
Observations from MeerKAT's L-band (900-1670 MHz) are widefield, with a single primary beam covering %$\sim 1 \degree$[squared] 
slightly more than one square degree (twice the area of the FIRST survey, from which RGZ takes its cutouts). At full resolution of $\sim$\SI{8}{\arcsecond} a single image will have close to 4K resolution in pixels and contain up to ten thousand sources depending on the observation depth \citep{heywood_mightee_2021}.
The other SKA-Mid precursor ASKAP, has similar specifications. In the era of SKA, it will be impractical to process observations down to individual source cutouts before applying machine learning methods to simplify data analysis. 

%high-impact motivation and results sentences
Even with the current state of the art, it is unfortunately still impractical to attempt to train vision networks with images as large as those produced by MeerKAT. Nevertheless, it is already important to explore the capabilities of SSL in order to extract generalized representations of larger, less-sparse, non-object-centric scientific images from the next generation of radio interferometers. Should such an approach succeed, it would be built on an easily scalable method for assembling datasets for SSL training. \AN{Furthermore, if the learned representations from such data are capable of transferring to both large field-of-view tasks such as source detection and small field-of-view tasks like morphology classification, the model can confidently serve the purpose of a foundation model. This emphasis on wide-field images for SSL training does not} negate the importance of Galaxy Zoo projects, which above all are useful because of labels provided by the efforts of large numbers of citizen scientists. Labeled datasets are essential for fine-tuning SSL-trained backbones to address specific tasks, as we demonstrate in this work.

\begin{figure}[h]
\includegraphics[width=.5\textwidth]{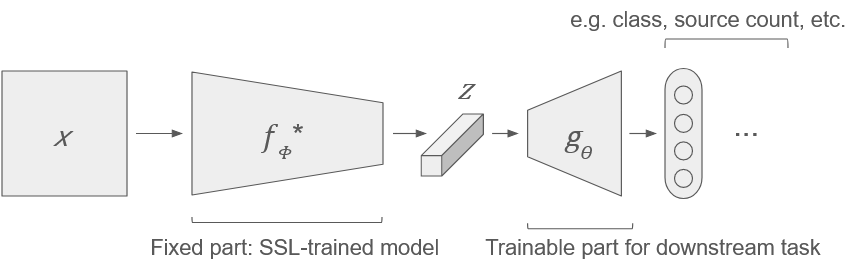}
\caption{Framework under investigation in this paper. Input data $x$ is fed into the SSL-trained network with fixed parameters, $f_{\Phi}^*$. Resulting embeddings $z$ are used by the trainable network $g_{\theta}$ to perform a specific downstream task.}
\label{fig:framework}
\end{figure}

The framework that we investigate is illustrated in Figure \ref{fig:framework}. The model $f$ with parameters $\Phi$ is trained in a self-supervised fashion. Input data $x$ is passed into this model with fixed parameters, $f_{\Phi}^*$, resulting in a vector representation of the latent space, otherwise known as embeddings $z$. These embeddings are used by a trainable network $g$ with parameters $\theta$, whose output is designed to address a specific downstream task, such as classification or regression. The primary focus of this work is to train the backbone model $f_{\Phi}$.

For our training data, we limit ourselves to only a fraction of publicly available MeerKAT observations --- images from the L-band continuum. We show that the DINO framework with both ResNet and Vision Transformer backbones is capable of encoding important information into a small latent space. Emerging properties indicate that it can separate sources from background, and evaluation on downstream tasks shows it can retain both major statistical properties of the data and individual source morphologies.

The structure of the paper is as follows. In Section \ref{sec:data} we describe the data used for both training and evaluation, as well as pre-processing we performed. In Section \ref{sec:method} we describe the DINO framework for self-supervised learning, the network training process, and emerging properties of the trained networks. Evaluation results are presented and discussed within the context of other foundation models in Section \ref{sec:evaluation}, before our concluding remarks in Section \ref{sec:summary}.

\begin{figure}[h]
\includegraphics[width=.5\textwidth]{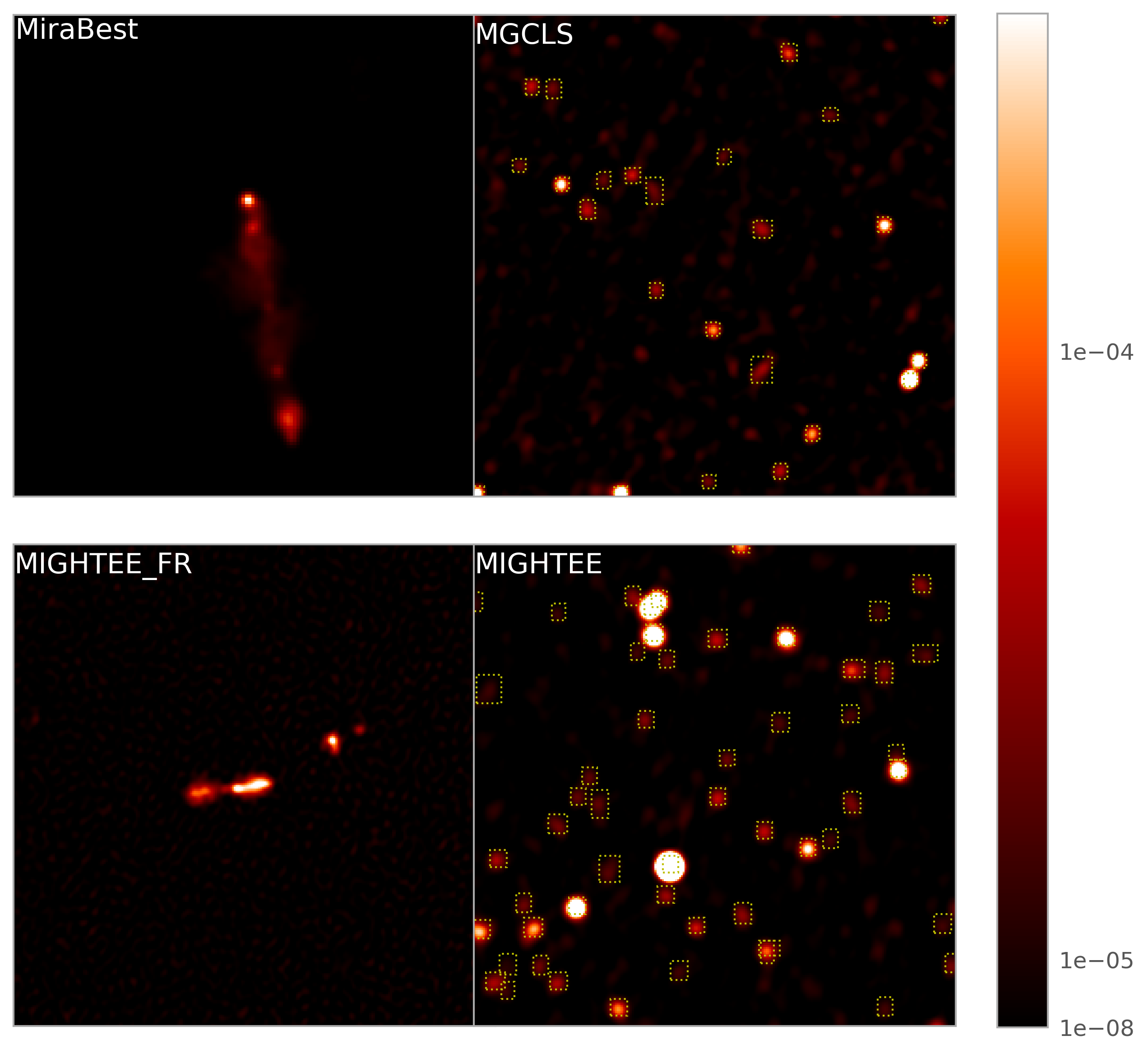}
\caption{Examples from datasets used in this work. Top left: a FRII galaxy from MiraBest (VLA). Bottom left: a FRII galaxy from MIGHTEE (MeerKAT). Top right: a 256x256 pixel cutout from Abell 209, a galaxy cluster observed as part of the MGCLS (MeerKAT). Bottom right: 256x256 pixel cutout from the COSMOS field, observed as part of MIGHTEE (MeerKAT). Yellow boxes indicate the extent of sources as given in the compact source catalogs. The color bar indicates the flux in mJy/beam for all images except MiraBest.}
\label{fig:sample_data}
\end{figure}

\section{Data}\label{sec:data}

To avoid confusion, we use the following terminology: a "field" is the target of a single MeerKAT observation, such as the COSMOS field or the galaxy cluster Abell 33; a "source" is a single (Gaussian or extended) radio source, usually a galaxy, in such a field; an "image" refers to the CLEAN reconstruction of the MeerKAT observed visibilities of an entire field, and a "crop" is a much smaller 2D cutout from such an image.

\subsection{MeerKAT}\label{sec:MeerKAT}

The SKA precursor MeerKAT is a radio interferometer composed of 64 dishes that observes the sky below a declination of +45\degree{} \citep{jonas_meerkat_2018}. The L-band (900 - 1670 MHz) system, corresponding to SKA-Mid, has a primary beam FWHM of 1.2\degree{} at 1.28 GHz. The array's layout features a dense core of antennas within a 1 km diameter and a maximum 7.7 km baseline, which enables high sensitivity to various angular scales. Full-resolution maps possess beam sizes of around 7.5 - \SI{8}{\arcsecond} and image noise levels of approximately 3 - 5 $\mu$Jy/beam, capturing extended structures up to tens of arcminutes.

\subsubsection{MGCLS}\label{sec:MGCLS}

The MeerKAT Galaxy Cluster Legacy Survey (MGCLS, \citet{knowles_meerkat_2022}) began in 2018, accumulating 1000 hours of L-band observations. 
The targets were 115 galaxy clusters between -80$\degree$ and 0$\degree$ in declination (DEC) spread out over the full range of right ascension (RA). % (see Figure \ref{fig:mgcls_sources}). 
\citet{knowles_meerkat_2022} calibrated and imaged the data, publicly providing CLEAN images and source catalogs in their first data release (DR1, \citet{sarao_meerkat_2021}). 

Each 8-12 hour observation was imaged by \nobreak{CLEANing} to a depth of $\sim$50 $\mu$Jy/beam, using robust weighting of -1.5 for the full resolution of $\sim7.5 - $\SI{8}{\arcsecond}. These 'basic' images were intended for visual inspection and source finding. Science-ready 'enhanced' images were then corrected for primary beam effects at each frequency; these are the images we use for constructing our dataset. The final Stokes-I continuum images are the inner 1.2$\degree \times 1.2\degree$ portion of the fields ($\sim$3500 $\times$ 3500 pixels but varies according to field), showing brightness at the reference frequency of 1.28 GHz. Observations that include very strong sources (I $>$ 100 mJy/beam) %(I $>\mathcal{O}(2)$ mJy/beam)
have limited dynamic range due to residual imaging artifacts. \AN{This and other data quality issues are discussed at length in \citet{knowles_meerkat_2022} Section 4.4.}

The survey team assigned data quality flags (DQF) as follows: 0 = good dynamic range; 1 = moderate dynamic range with some artefacts around bright sources; 2 = poor dynamic range with high contamination by bright source artefacts; 3 = poor dynamic range with ripples across image. \AN{Examples of images with each DQF are shown in Figure \ref{fig:dqf}.}

\subsection{Datasets}\label{sec:datasets}\

\begin{table*}
\begin{tabular}{p{0.15\paperwidth} p{0.33\paperwidth} p{0.08\paperwidth} p{0.09\paperwidth}p{0.09\paperwidth}}%[width=\pagewidth]
\textbf{Name} & \textbf{Origin} & \textbf{Purpose} & \textbf{Total crops} & \textbf{Crop shape} \\
\hline
MGCLS\_20k   & MGCLS enhanced images\tablefootnote{https://doi.org/10.48479/7epd-w356}   &  Training     & 19554 & 256$\times$256 \\
MGCLS\_5k   & MGCLS enhanced images   &  Training     & 5000 & 256$\times$256\\
%MGCLS\_offset & MGCLS basic images & Evaluation & 5000\\
%MIGHTEE   & MIGHTEE COSMOS and XMMLSS images\footnote{CLEAN with Briggs robust weighting -1.2}     &  Evaluation     & 3K     \\
MIGHTEE  &  MIGHTEE COSMOS and XMMLSS images\tablefootnote{https://www.mighteesurvey.org/data-access}  &  Evaluation     & 1246     & 256$\times$256\\
MIGHTEE\_FR  &  FRI/FRII galaxies in MIGHTEE fields  &  Evaluation     & 174  & 150$\times$150   \\
MiraBest\tablefootnote{https://zenodo.org/records/4288837}  &  VLA NVSS and FIRST survey images  &  Evaluation     & 1256     & 150$\times$150\\
RadioGalaxyDataset\tablefootnote{https://github.com/floriangriese/RadioGalaxyDataset}   &  VLA FIRST survey images  &  Evaluation     & 2158     & 300$\times$300\\
%Astronomaly   & MGCLS enhanced images     &  Anomaly detection     & 9582 \\                      
\hline
\end{tabular}\caption{Datasets used in this work. \AN{The first three consist of random multi-source crops from wide-field images, while the last three consist of curated postage-stamp cutouts of individual galaxies.}}\label{tbl:datasets}
\end{table*}

While MGCLS data is used for training, different datasets are used for evaluation of various tasks. The datasets are summarized in Table \ref{tbl:datasets}. Sample images are shown in Figure \ref{fig:sample_data}. All the continuum images are centered at the corresponding observing band and thus contain only one "channel" (in comparison to typical three-channel RGB images), containing values typically ranging from $-10^{-10}$ to $10^{-3}$ mJy/beam. 

\subsubsection{Datasets for training}\label{sec:training_data}

We train the network on the MGCLS enhanced images, since the more challenging task of training on the basic images can wait until the SSL method is known to perform well on science-ready data. Similarly, we do not incorporate MIGHTEE crops into the training dataset, to avoid complications from observations of very different depth and sensitivity.

To make the wide-field MeerKAT images tractable for deep learning networks, images are subdivided into crops. In the field of computer vision, networks are often trained on small images, and for DINO, our deep learning framework of choice, input is by default randomly cropped to a size of 224 $\times$ 224 pixels. We choose a slightly larger crop size of 256 $\times$ 256 pixels, which allows for some slight variation in how the final crop is performed each time the data batch is loaded. Overlapping areas between adjacent crops are at maximum ten pixels on a side, since first we find the closest integer number of crops that fit in an image, and then arrange the crops such that overlap is minimized. 
No images are excluded from the training based on the data quality flag. Therefore, the full training dataset \textit{MGCLS\_20k} contains 19554 crops from the MGCLS DR1 dataset. A randomly selected 5000-crop subset named \textit{MGCLS\_5k}, containing at least one crop from each field, is used for hyperparameter optimization as well as for training the memory-intensive Vision Transformer backbone, whose attention mechanism is computationally more expensive than the convolutional operations that ResNet uses.  %[as well as to test the effectiveness of fine-tuning from ImageNet weights as opposed to training from scratch with much more data.] 

\subsubsection{Datasets for evaluation}\label{sec:eval_data}

To evaluate network performance on compact source count prediction, we use portions of the MGCLS dataset as well as crops from the MIGHTEE survey \citep{heywood_mightee_2022}. Datasets used for the downstream task of galaxy morphology classification are the MiraBest dataset \citep{porter_mirabest_2023}, 174 hand-labelled sources from the MIGHTEE survey \AN{\citep{
scaife_mightee_2023}}, and \citet{griese_floriangrieseradiogalaxydataset_2022}'s FIRST dataset. 

The MeerKAT International GHz Tiered Extragalactic Explorations (MIGHTEE) survey was designed to reach the confusion limit in the total intensity continuum \citep{Jarvis:20184F, heywood_mightee_2022}. It is a deeper L-band survey than MGCLS, reaching a sensitivity of $\rm 2~\mu Jy~beam^{-1}$ via $\sim$17 hours of on-source time. %As a deeper survey, $\sim10000$ sources were detected in the single-pointing COSMOS image alone. %c.f. 3k? in a MGCLS pointing
The Early Science data release \citep{heywood_mightee_2021} contains continuum data products for two of the four observed fields: COSMOS and XMMLSS (XMM-Newton Large Scale Structure field). We take 256 $\times$ 256 pixel crops from these continuum images in the same way as for MGCLS, described in the previous section. These crops are taken from the Early Science images with a similar CLEAN weighting (robust -1.2) and the same resolution to MGCLS (\SI{8.6}{\arcsec} and \SI{8.2}{\arcsec} for COSMOS and XMMLSS respectively). For the morphology classification dataset \textit{MIGHTEE\_FR}, we simply make crops centered on the coordinates of the labeled galaxy.

MiraBest is a publicly available dataset manually labelled according to Fanaroff-Riley morphological classification \citep{fanaroff_morphology_1974}, sourcing its data from the Very Large Array's (VLA) NVSS and FIRST surveys. Because of this, there is some overlap between MiraBest and Radio Galaxy Zoo. Morphological labels are accompanied by a confidence parameter of either "confident" or "uncertain". Crops have an angular scale of \SI{270}{\arcsecond}. In the literature, it is common to use only the 729 "confident" samples. %if we match the scale with mightee, mention it here

\citet{griese_floriangrieseradiogalaxydataset_2022}'s RadioGalaxyDataset is also from FIRST survey images, but includes two additional morphology classes: compact sources and bent-tailed galaxies.% Crops are larger, at 300 $\times$ 300 pixels. %mention the class balance/imbalance of these datasets?

%Because of the different positions of the MGCLS target fields on the sky, the basic image sizes vary slightly from field to field. The number of 256x256 crops that could be taken from each image were therefore calculated based on fitting an integer number of crops across the image width (identical to the height) with a minimal amount of overlap between the crops. This ranged from a total overlapping pixels/image to b total overlapping pixels/image. Once the crop coordinates had been determined for the basic images, they were used to crop the corresponding region from the enhanced images. Due to [reasons, but not including offset applied, mainly primary beam correction i would say],  
%MGCLS paired dataset ^^

%, with overlap determined based on taking an integer number of crops across the image width/height and ranging from a to b pixels (or percentages)
%more details of paired dataset; downscaling of enhanced image to match if needed

\subsection{Source Catalog Labels}\label{sec:labels}

Although images from the MGCLS and MIGHTEE surveys are unlabeled, certain useful characteristics can be inferred from the observation metadata and the extracted source catalogs. These can be treated as "noisy" labels, in that there is a certain degree of uncertainty associated with each label. Of key interest are the quantities calculated from the compact source catalog (a sample is given in Table 2 of \citet{knowles_meerkat_2022}): the number of sources per crop, their shapes, locations, and fluxes. For visualization purposes, metadata from both the FITS image headers and Table 1 of \citet{knowles_meerkat_2022} are useful, although observation metadata describes the entire field rather than each individual crop. Two fields have additional labels available: Abell 209 and Abell S295 had an optical cross-match performed, and extended sources were identified by experts. 

Both the MGCLS compact source catalog and the MIGHTEE catalogs were obtained via PyBDSF \citep{mohan_pybdsf_2015}, using the default parameters (island and source detection thresholds ($3\sigma_{local}$ and $5\sigma_{local}$ respectively). Source detection on MGCLS was performed on CLEAN images with a robust weighting of -1.5, which is close to uniform weighting (robust -2), resulting in more system noise. MIGHTEE source detection was performed on the image with a robust of 0, which is an intermediate weighting scheme between natural and uniform weighting. MGCLS excluded sources detected within a certain radius of ultra-bright sources, while for the MIGHTEE survey such exclusion was based on visual inspection. 

Referring to Figure \ref{fig:sample_data}, we can see some slight differences in the resulting source catalogs. MIGHTEE has detected some sources that overlap, shown in the upper, left-of-center part of that crop. This is not an uncommon occurrence in the MIGHTEE source catalog; in fact, extended sources such as the FRII galaxy shown in the lower left of the figure would often have each component of this multi-Gaussian source included in the catalog. Multi-Gaussian components are excluded entirely from the MGCLS catalog, but not from MIGHTEE. Additionally, while even very faint areas of the MIGHTEE crop are designated sources, this does not appear to be the case for the MGCLS crop, where several faint blobs in the bottom central part of the image are not designated sources while one similar-looking one is. %is this too much detail?

PyBDSF has been shown to reach both purity and completeness of at least 90\% for a signal-to-noise ratio (SNR) of source peak flux over image rms flux $\geq$ 4.3 using simulated, naturally weighted MeerKAT images \citep{vafaeisadr_deepsource_2019}. The same study suggests that, even on images consisting solely of point sources, completeness only approaches 100\% near SNR $=$ 6. Hence we consider the source catalog entries as noisy labels, rather than ground truth, which for real data such as ours is additionally complicated by the presence of extended and overlapping sources.

\subsection{Image Pre-processing}\label{sec:pre-processing}

\begin{figure}
\includegraphics[width=\columnwidth]{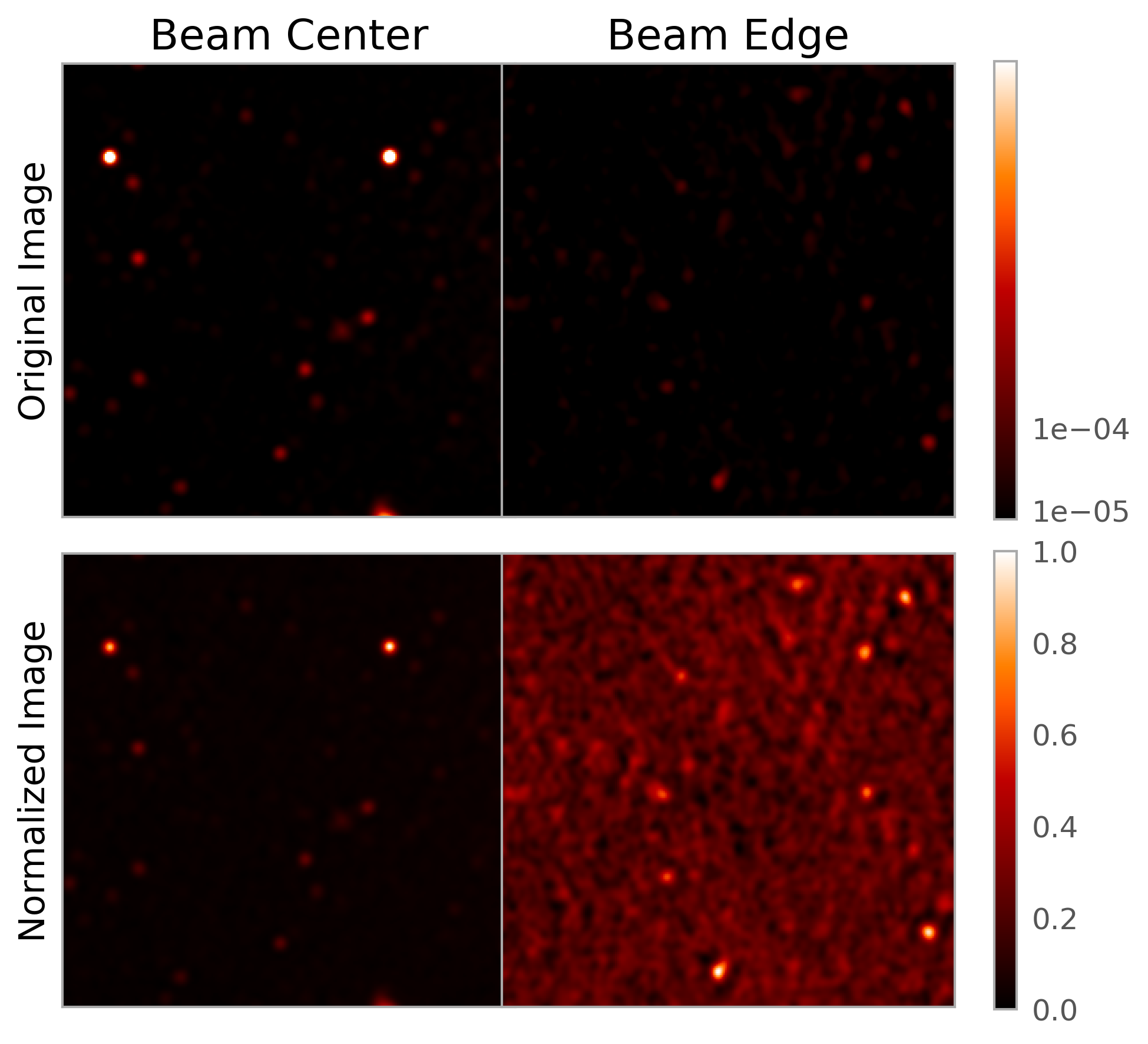}
\caption{The relative noise level and source brightness varies according to location in the primary beam, shown in this example from the MGCLS field Abell 209. Simple per-crop normalization distorts the perception of relative brightness of the sources.}
\label{fig:rawdata}
\end{figure}

\begin{figure}
\includegraphics[width=\columnwidth]{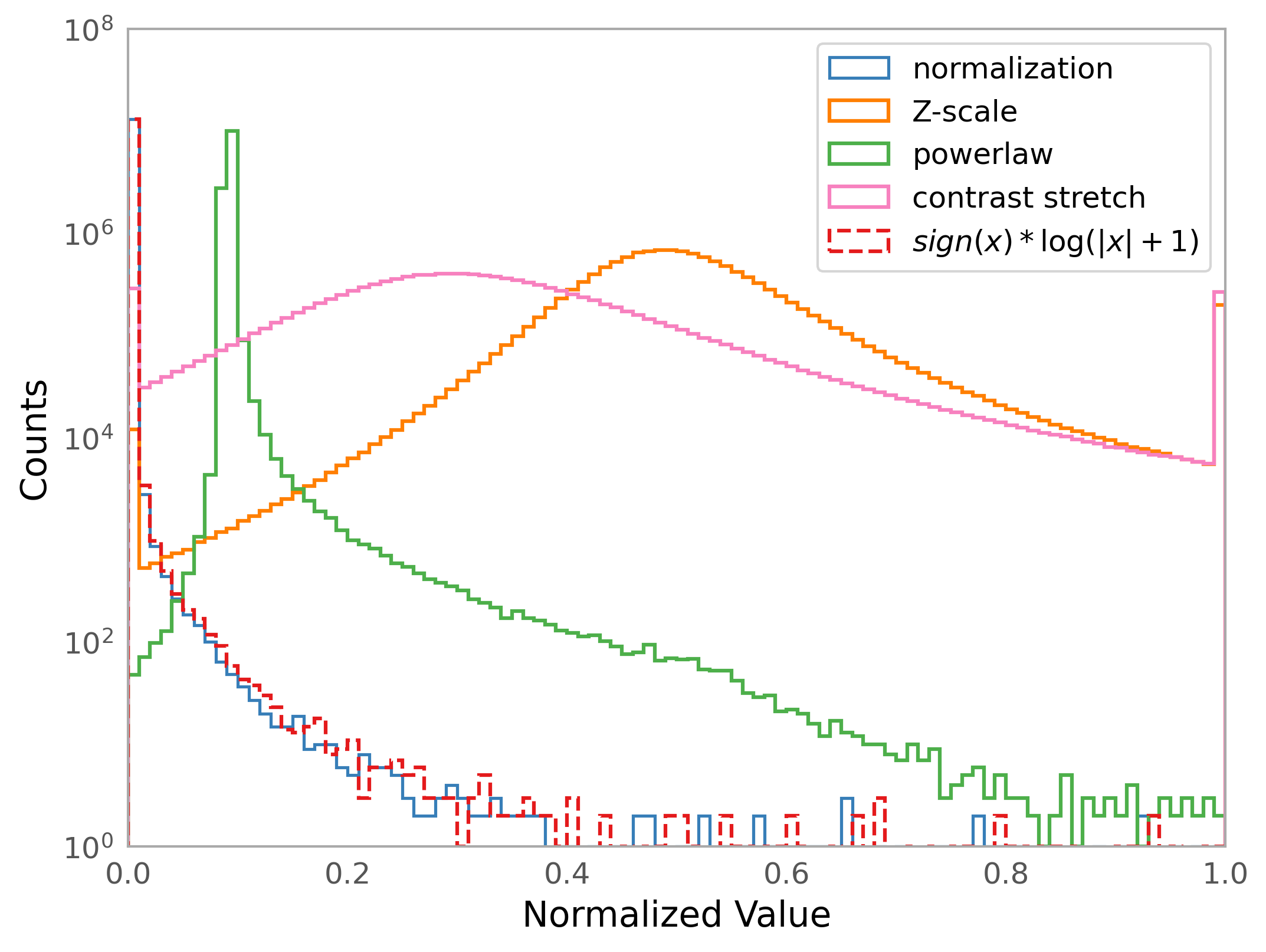}
\caption{Various methods for scaling high dynamic range images for use with deep learning networks, applied to the image of Abell 209. Powerlaw scaling is done with a factor of $\frac{1}{10}$, and contrast stretching is performed with the 2nd and 98th percentiles.}
\label{fig:preprocessing}
\end{figure}

A common problem when preparing scientific images for machine learning is correct scaling of the images. This is especially relevant for images which are not centered on a single, brightest source as are Galaxy Zoo images, but ones which may contain numerous sources of different magnitudes. The flux values in each crop could simply be normalized as is done in Galaxy Zoo; however, this removes certain context information. 

Figure \ref{fig:rawdata} illustrates the difference for 256$\times$256 pixel crops near the center of the primary beam (left images) and near the edge (right images). There is about an order of magnitude difference between the brightest sources in the images on the left and the right, a characteristic that is not preserved by simple normalization. Additionally, the increase in noise near the edge of MeerKAT's 1.2$\degree$ primary beam is greatly exaggerated by such scaling.
As we prepare for the full scale of SKA data we must assume that future deep learning techniques will be able to overcome the problem of images with more than 4000 pixels on a side, and possibly even neglect images altogether in favor of visibility data. Therefore we choose to apply scaling based on the image of each field rather than for each crop individually. 

The distribution of noise and sources in an image is skewed towards the value of the noise, as sources are considerably more rare than background. This may not be the case in some individual crops, such as those containing a large portion of an extended source, but those are rare in this dataset. \citet{knowles_meerkat_2022} estimate that the number of extended sources in each field is 3-5\% of the number of compact sources, and of those, the angular extent required to occupy most of a crop would be 0.6$\degree$. Therefore, even for a single crop the flux brightness histogram should be strongly skewed towards the rms, with a long (not necessarily continuous) tail on the right indicating the presence of sources of various orders of magnitude. A good scaling technique should still have most counts centered around the image rms, while emphasizing the non-noise values of the sources.

Figure \ref{fig:preprocessing} compares several approaches to image scaling, utilizing different techniques from the fields of computer vision, astronomy, and particle physics. Note that, due to the size of the full image of a single field, the y-axis is displayed in log scale.
The first method, z-scaling, originates from the need to calculate histograms based on a low number of samples, due to computing limitations. It is commonly employed in astronomy, and was used by \citet{riggi_astronomical_2023} for the task of object detection, although they speculated that contrast stretching might have been more appropriate instead. 
Indeed, we see that contrast stretching (e.g. \citet{negi_hybrid_2014}), a technique often used in computer vision to re-distribute values between a low and high percentile (generally 2\% and 98\%), results in a more desirable distribution compared to z-scaling; the majority of values are concentrated at a lower value, allowing more space for the long tail. Both these methods cause the brightest sources to be assigned the same maximum value, but the occurrence of such bright pixels can be seen to be extremely small through comparison with the histogram of normalized values.  %it does this re-distribution pretty much by clipping and then regular normalization though?

Another method often used in cosmology and particle physics, scaling the data by $sign (x) * \log(|x| + 1)$ where $x$ is the data vector or array, performs similarly to standard normalization, while adding some emphasis on the brighter pixels. For our data, the resulting distribution could be improved by replacing the very brightest pixels with a fixed value. However, the advantage of this method - being easily invertible without having to store the values of the low and high percentiles - would then be lost.
Power law scaling is also useful in particle physics, accomplished by dividing by the maximum and then scaling to some power, usually $\frac{1}{10}$ as shown here \citep{drozdova_radio-astronomical_2023}. Similar to normalization, this causes the distribution to be dominated by the noise, although the scaling of the brighter pixels is much improved. 

For this work, we chose to scale images with contrast stretching, since it results in a distribution that emphasizes the sources, can still be inverted, and is familiar to the computer vision community. MGCLS images were contrast stretched using the 2nd and 98th percentiles, while MIGHTEE images used the 2nd and 99.9th percentiles, due to the presence of a few spurious bright pixels. 

\section{Method}\label{sec:method}

\subsection{DINO}\label{sec:dino}

DINO, which stands for knowledge distillation with no labels, is a SSL framework \citep{caron_emerging_2021} that trains a student network to match the output of a teacher network. Like with BYOL, it does not use negative pairs; DINO inputs different views of an image into each network, rather than into a single encoder network. %The output of the student network is then trained to match as closely as possible to that of the teacher network's predictions, using cross-entropy loss $H$, where $H(a,b) = -a \log b$. %(see Figure \ref{fig:dino}). 
The student and teacher networks must have matching architectures, although this can take different forms, notably allowing use of Vision Transformers (ViT) as well as standard CNNs such as ResNet.

Using the standard DINO configuration, two global views, $x_1^g,$ and $x_2^g$, are passed to the teacher network, while all global and local views $V$ go to the student network. This design encourages the network to discover local-to-global correspondences. The loss across all views is as follows:

\begin{equation}
   \underset{\theta_s}{min} \underset{x \in \{x_1^g,x_2^g\}}{\sum} \underset{\underset{x'\ne x}{x' \in V}}{\sum} H(P_t(x),P_s(x')),
\end{equation}\label{eq:DINO_loss}

where $\theta_s$ are the parameters of the student network, and $P_t$ and $P_s$ are the network output probability distributions for the teacher and student networks respectively, and the cross-entropy loss $H(a,b) = -a \log b$.

Parameters of the student network $\theta_s$ are learned using stochastic gradient descent. The teacher network's parameters $\theta_t$ are updated using a momentum encoder, which acts as a form of model ensembling, allowing the teacher's better performance to guide the training of the student. %if necessary, mention normalization used to avoid collapse 

The network that we refer to as the SSL-trained backbone model $f_{\Phi}$ as in Figure \ref{fig:framework} is the teacher network.

% \begin{figure}
% \includegraphics[width=0.5\textwidth]{dino_diagram_paper.png}
% \caption{DINO diagram, from \citet{caron_emerging_2021}.}
% \label{fig:dino}
% \end{figure}

\subsubsection{Data Augmentations}\label{sec:augmentations}

DINO depends on data augmentations to distill knowledge from different parts of the input images. 
%for the astronomer, what does this mean?
Data augmentation is commonly used in deep learning to build a more robust network, by distorting the data in ways likely to occur in the environment in which the data is collected, thereby enriching the training dataset.
At the very basic level are the local and global views passed to the networks. Global views are large random crops of size 224$\times$224 pixels, randomly re-scaled between 0.4 and 1.
Local views are smaller 96$\times$96 pixel crops from random areas of the input image, which also vary randomly in scale. In our work, we chose to use eight local crops and kept the default suggested scaling ranges of 0.05 - 0.14 for ResNet and 0.05 - 0.4 for ViT. Additional augmentations are applied to both local and global views in order to improve the robustness of the network.

The standard DINO image augmentations are optimized for natural images, and consist of a random horizontal flip, random color jitter, Gaussian blur, and solarization. Because MGCLS continuum images are single channel, we chose not to use color jitter and solarization, which are designed for RGB images. Other fields of physics have had great success with science-based augmentations (e.g. \citet{strong_impact_2020}), which are slowly starting to be implemented for radio astronomy. 

Previously, \citet{slijepcevic_learning_2022} found that random resized cropping was the augmentation with most significant impact to BYOL performance, with rotation as the second most important. Random cropping at various scales is already a core part of creating DINO's local crops, so we evaluate the effects of only random rotation, random autocontrast, and random power law scaling. For our non-object-centric data, with fluxes decreasing near the edge of the primary beam as shown in Figure \ref{fig:rawdata}, the latter two augmentations may improve the ability of the network to distinguish source from noise in low-contrast areas of the field.

We test the effectiveness of rotation, autocontrast, and power law scaling by adding one of these random augmentations at a time in addition to the standard ones of flipping and Gaussian blur, before trying different combinations.
Using a ResNet50 backbone, we train for 100 epochs on the smaller \textit{MGCLS\_5k} dataset, and consider both the final training loss and the mean squared error (MSE) of our linear evaluation task (see Section \ref{sec:linear}) in determining the impact of additional augmentations. In this way, we find that the combination of rotation and power-law scaling gives the best results, performing better on the linear evaluation task than a network trained using the default augmentations, while having a similar final training loss. %.with a final MSE of 147 vs 150 for the default augmentations and similar training losses of 10.55 vs 10.35.

\subsection{Training}\label{sec:training}

We train backbones of two different architectures: a convolutional ResNet and a Vision Transformer. Because weights for a ResNet50 network pre-trained via DINO with ImageNet are publicly available, we use this same network depth.
For the ViT backbone, we use the small 21-million parameter network with the DeiT implementation \textit{deit\_small} \citep{touvron_training_2020}. We use a patch size of 8, as this corresponds more closely to the average size of a compact source in the MGCLS catalog than the default patch size of 16. No pre-trained weights for this configuration were available, so we initialize both backbones with random weights.

We perform a small grid search to select certain hyperparameters: the learning rate, batch size, and momentum parameter for the teacher network. Recall that the teacher network updates via a momentum encoder, performing an exponential moving average on the weights of the student network. For training a ResNet, a constant weight decay was recommended, so we only include variations of this hyperparameter in the search for the ViT backbone, where memory limitations constrained the batch size of the ViT backbone to 8. In each training epoch, the network is exposed to the entire dataset, no matter the batch size. Networks are trained for 100 epochs on the \textit{MGCLS\_5k} dataset, and then evaluated on the linear regression task. Hyperparameters which minimize both the training loss and MSE are listed in Table \ref{tbl:params} along with other network parameters.

\begin{table}[h]
\textbf{ResNet}
\begin{tabular}{p{0.4\columnwidth} p{0.3\columnwidth}}%[width=\pagewidth]
Parameter & Value \\
\hline
architecture   & resnet50   \\
batch size per GPU   & 32   \\
learning rate   & 0.03   \\ %base learning rate? scaled by 256
momentum   & 0.996  \\
weight decay   & 1e-4    \\
& \\
\end{tabular}
\textbf{ViT}\hspace{1cm}
\begin{tabular}{p{0.4\columnwidth} p{0.3\columnwidth}}%[width=\pagewidth]
Parameter & Value \\
\hline
architecture   & deit\_small   \\
patch size   & 8   \\
batch size per GPU   & 8   \\
learning rate   & 0.00625  \\ %base learning rate? scaled by 256
momentum   & 0.996  \\
weight decay   & 0.04     
\end{tabular}
\caption{DINO backbone parameters. During training, the learning rate is scaled by the factor $\frac{batch\_size}{256}$.}\label{tbl:params}
\end{table}

\begin{figure}
\includegraphics[width=0.5\textwidth]{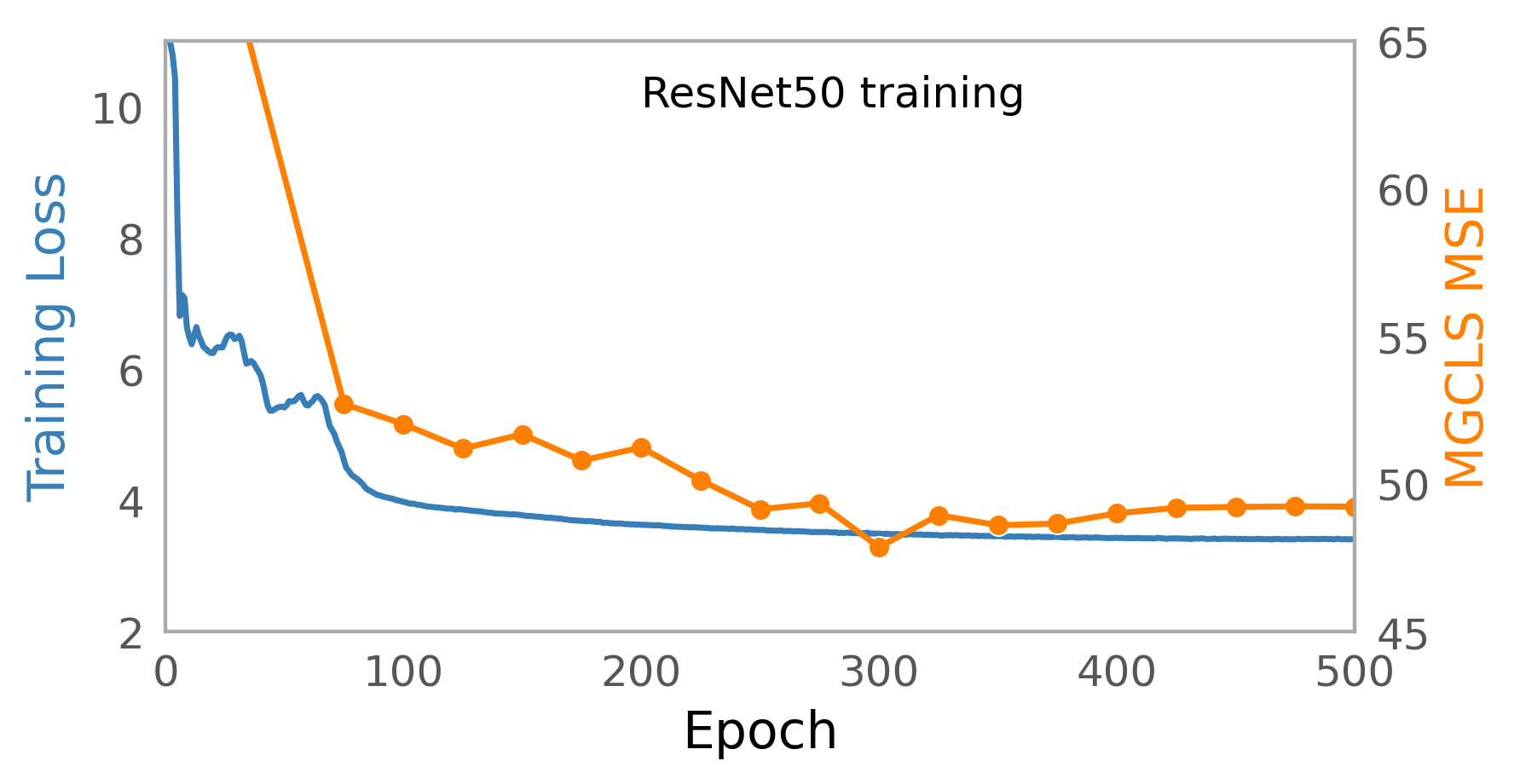}
\includegraphics[width=0.5\textwidth]{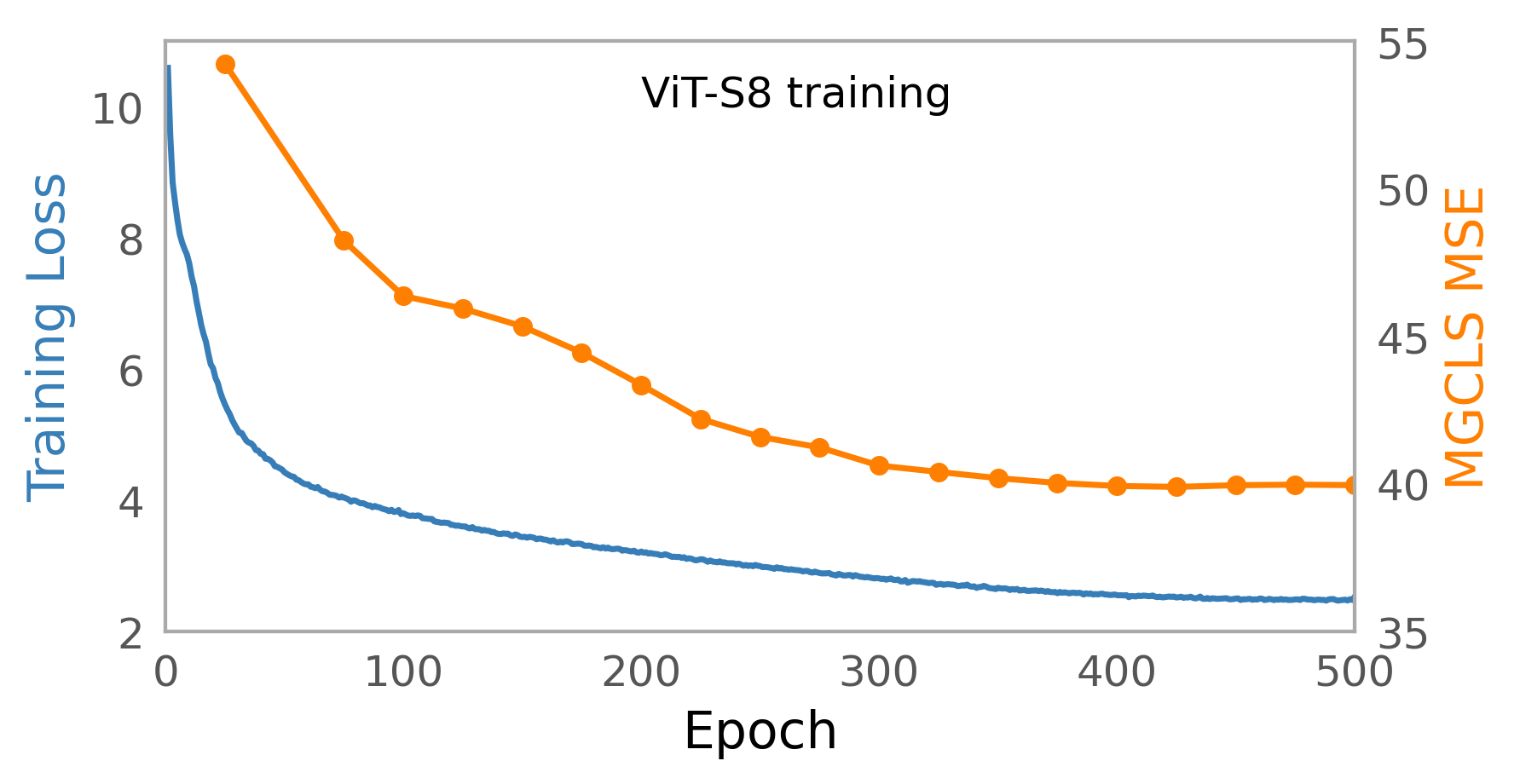}
\caption{Training loss and source count linear regression (see section \ref{sec:linear} for details) MSE per evaluation epoch, for both backbones.}
\label{fig:mgcls_train}
\end{figure}

Our Resnet backbone is trained with the \AN{training split of the} \textit{MGCLS\_20k} dataset simultaneously on 6 GPUs, which are either NVIDIA TITAN X or Tesla P100, depending on availability. We train a ViT backbone on the same cluster but using the smaller \AN{training} dataset \textit{MGCLS\_5k}. The networks are evaluated every 25 epochs by performing linear regression to the compact source count. This is the only evaluation task that uses data from MGCLS rather than a different survey, so it is suitable as an initial evaluation task. \AN{Training of the linear regression is performed on a 70\% split of the original training dataset, and validated with the remaining 30\%. Finally, the MSE reported is that of the test set, which is a 20\% split from the \textit{MGCLS\_20k} dataset that does not overlap with any samples in \textit{MGCLS\_5k}.} 

We consider the training loss to have converged if either the loss did not decrease by more than 1\% in 100 epochs, or the source count linear regression MSE did not improve in the same period. Under this condition, we consider the 23-million parameter ResNet sufficiently trained after 425 epochs. %, which took four days and six hours. 
Because it is trained with a smaller dataset, training loss for the ViT-small backbone keeps decreasing \AN{by more than 1\%. Therefore, once the MSE no longer decreases, we stop training.}
The 21-million parameter ViT-small backbone converges at epoch 475 (see Figure \ref{fig:mgcls_train}). %after [24 hours] of training.

\subsection{Emerging Properties}\label{sec:emerging}

\begin{figure}
\centering{\includegraphics[width=.9\columnwidth]{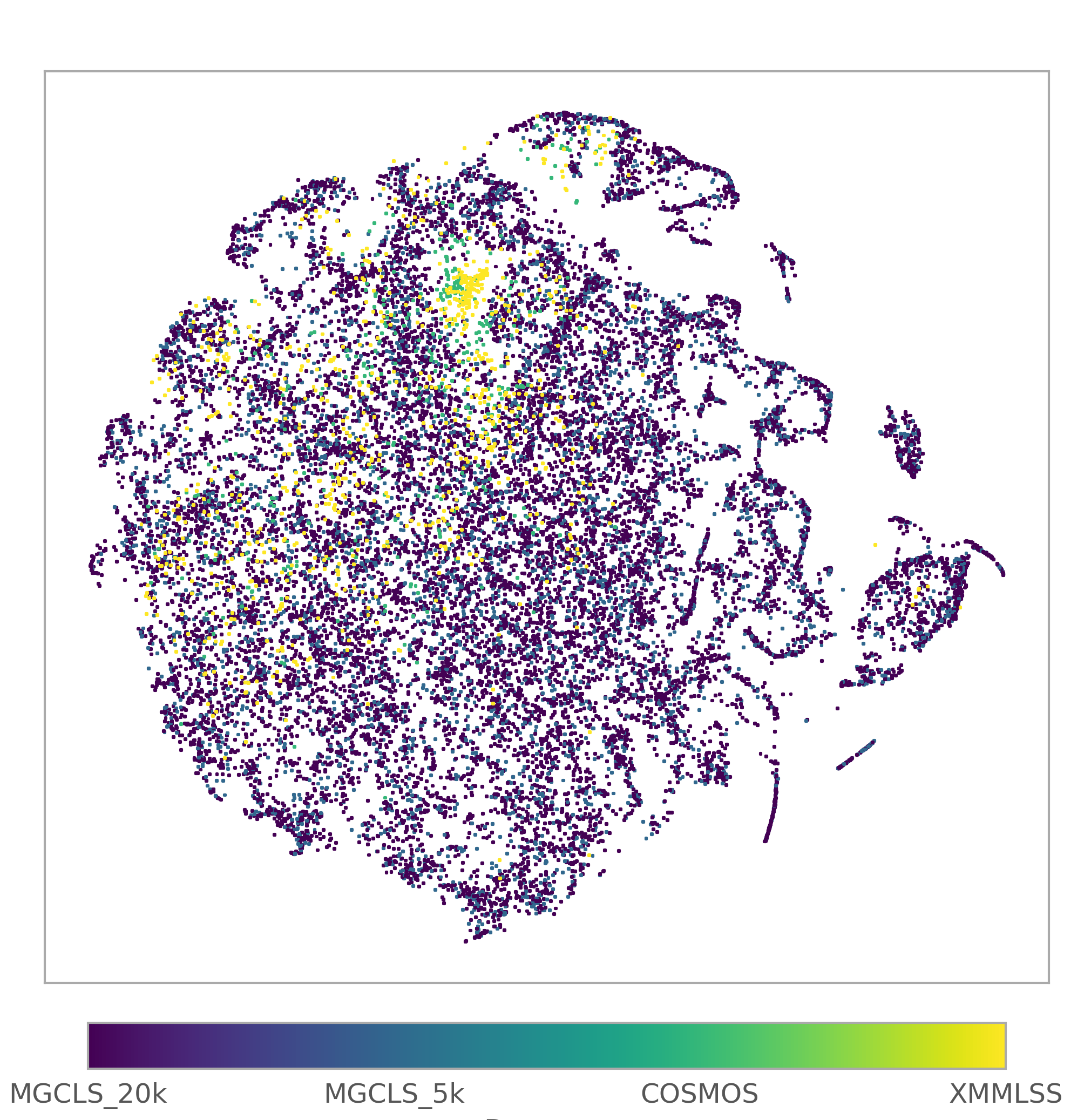}
\includegraphics[width=.9\columnwidth]{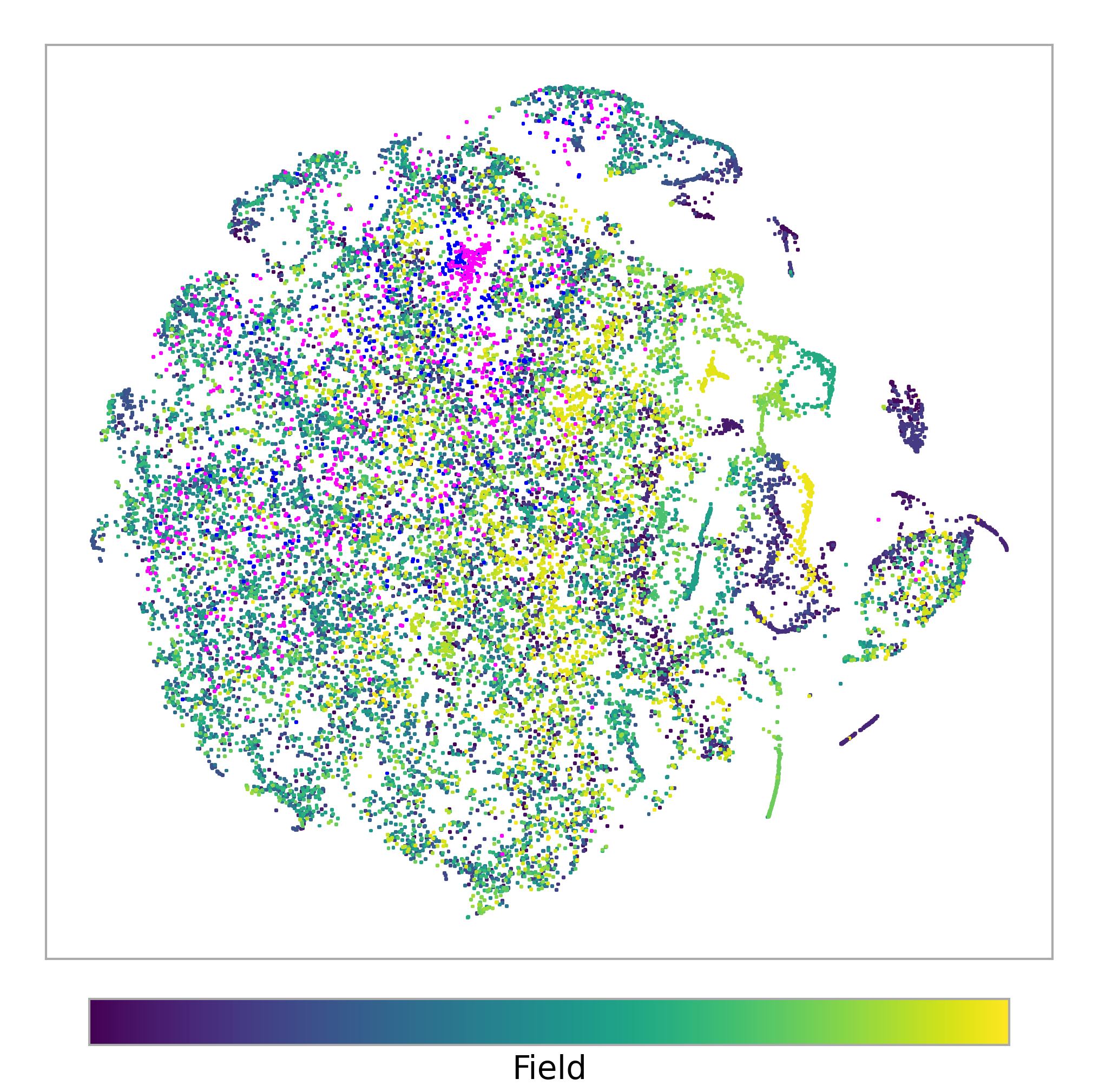}
\includegraphics[width=.9\columnwidth]{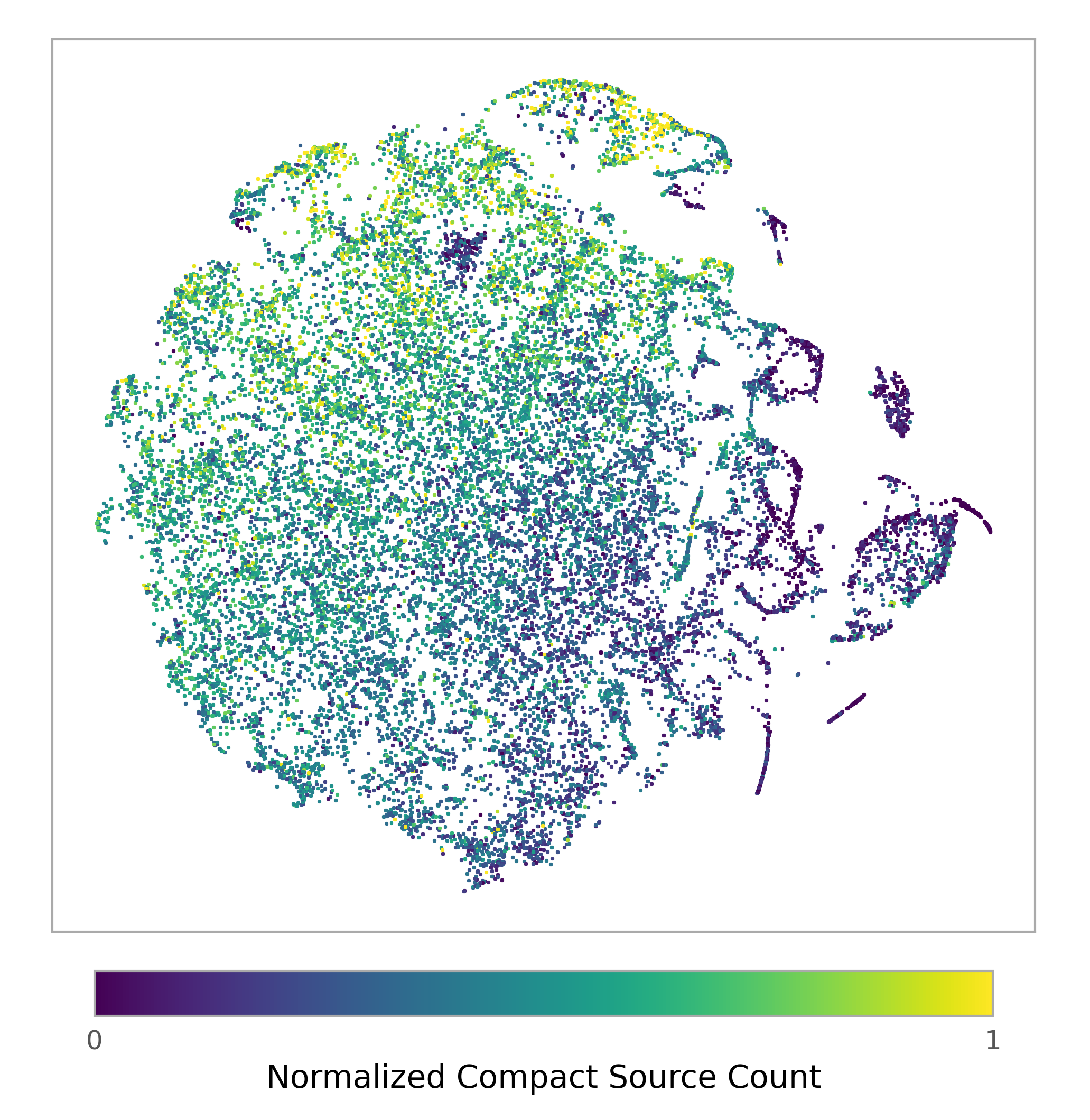}}
\caption{Features extracted from our trained Resnet backbone, visualized via t-SNE in two dimensions. The top panel shows patterns by dataset; the middle by observation field, with COSMOS and XMMLSS in blue and magenta, respectively; and finally by the number of compact sources per crop.}
\label{fig:mgcls_feats}
\end{figure}

To visualize emerging properties of the trained backbones, we employ the common dimensionality-reduction algorithm t-SNE and take advantage of the Vision Transformer architecture to show attention maps. 

Figure \ref{fig:mgcls_feats} shows the features of both the \textit{MGCLS\_20k} and \textit{MIGHTEE} datasets, extracted from our trained ResNet model, represented in two dimensions instead of the 2048 of the embedding layer. Different color-codings demonstrate how after training, the features are structured according to the dataset (top figure), observed field (middle), and number of compact sources per crop (bottom). It is clear that parts of the \textit{MIGHTEE} dataset (COSMOS field in blue, XMMLSS field in magenta) are distinct from the MGCLS data; as a deeper, more sensitive survey this is unsurprising. Likewise, certain fields in the \textit{MGCLS\_20k} dataset are more unique than others (the group in light green on the bottom-right corner are all crops from the field J1248.7-4118, which has a DQF of 2, indicating contamination by bright source artefacts). \textit{MGCLS\_20k} itself is heterogeneous even without the addition of \textit{MIGHTEE}, but even so the network can capture global properties of the individual crops, as evidenced by the last panel, showing the rough grouping of feature vectors according to the number of compact sources per crop. 

\begin{figure}
\includegraphics[width=\columnwidth]{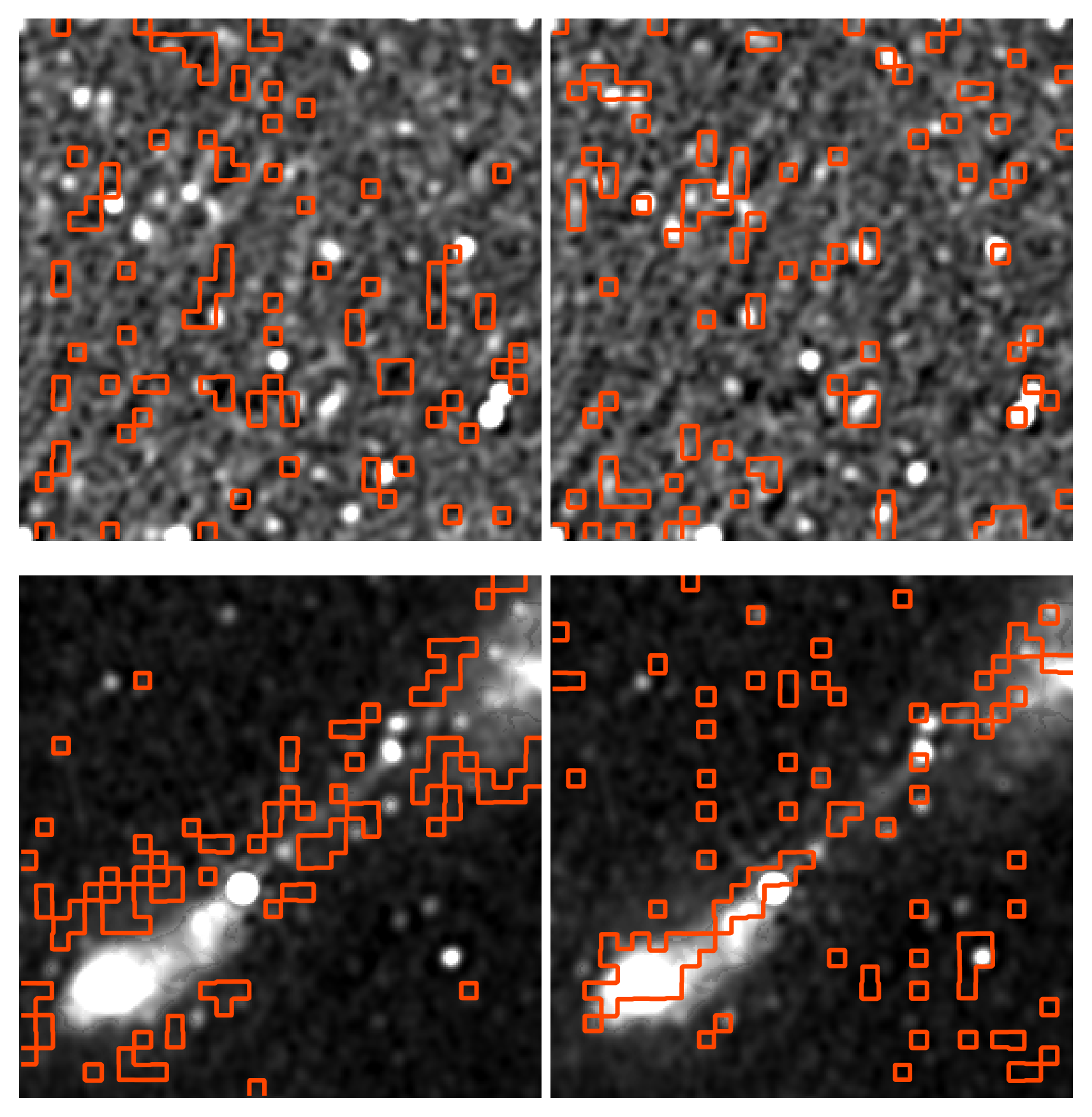}
\caption{Attention maps overlaid on the crop of Abell 209 from Figure \ref{fig:sample_data} (top) and an extended source from the same field (bottom). The left and right panels show maps from two of ViT-small's six attention heads. As the network architecture has a patch size of eight, the 90th percentile of the attention maps shows the 8$\times$8 patches that the specific attention head considers most important in the image.}
\label{fig:attention}
\end{figure}

While the t-SNE visualization can illustrate some global trends the model has learned, the details are revealed by visualizing the attention layers in the ViT network. Figure \ref{fig:attention} illustrates this for two crops from the field Abell 209 - the top example is the same crop shown in Figure \ref{fig:sample_data}. Red contours show the 8x8 patches in the attention heads at the 90th percentile. Such pixels highlight areas in the image where a particular part of the network is focusing its attention. 

In both these examples --- one of a typical crop and another containing a large, bright, extended source --- different attention heads pick out different attributes of the crops. Bright pixels or groups of pixels are more important to the attention heads visualized on the right side of the figure; darker areas, or areas bordering bright ones, are what is emphasized on the images on the left.

Although these visualizations are only a qualitative way of assessing the trained models, they give us some indication of the features learned by the networks.

\section{Evaluation and Discussion}\label{sec:evaluation}

\begin{table*}
\begin{tabular}{@{}p{0.14\paperwidth} p{0.23\paperwidth} p{0.18\paperwidth} p{0.07\paperwidth}p{0.08\paperwidth}p{0.08\paperwidth}}
%{@{}p{0.12\pagewidth} p{0.2\pagewidth} p{0.15\pagewidth} p{0.08\pagewidth}p{0.08\pagewidth}p{0.1\pagewidth}}%[width=\pagewidth]
\textbf{Backbone Name} & \textbf{Source} & \textbf{Training Dataset} & \textbf{Training Samples} & \textbf{Training Method} & \textbf{Backbone Network} \\
\hline
\textbf{MGCLS Resnet}   & \textbf{this work}   &  \textit{MGCLS\_20k}  & 20 K   & DINO & Resnet50 \\
\textbf{MGCLS ViTS}   & \textbf{this work}   &  \textit{MGCLS\_5k} & 5 K    & DINO & ViT-S8 \\
RGZ BYOL  &  \citet{slijepcevic_radio_2024}  &  Radio Galaxy Zoo & 108 K     & BYOL     & Resnet18 \\
GZ2 MoCo  &  \citet{hayat_self-supervised_2021}  &  Galaxy Zoo 2 & 1.2 M   & MoCo v2  & Resnet50   \\
DINO Resnet  &  \citet{caron_emerging_2021}  &  ImageNet & 1.3 M   & DINO     & Resnet50 \\
DINOv2 ViTG  &  \cite{oquab_dinov2_2024}  &  DINOv2 & 143 M   & DINOv2     & ViT-G14 \\
DeCALS BYOL  &  \citet{mohale_enabling_2024}  &  Galaxy Zoo DeCALS & 270 K    & BYOL     & Resnet18 \\
RGZ VDVAE  &  \cite{andrianomena_radio_2024}  &  Radio Galaxy Zoo & 108 K  & VDVAE     & Resnet34 \\
%Astronomaly   & MGCLS enhanced images     &  Anomaly detection     & 9582 \\                      
\hline
\end{tabular}\caption{Backbones referenced in this work. \AN{All of these backbones are trained in a self-supervised manner, even if they are trained on different datasets and with different network architectures.} }\label{tbl:backbones}
\end{table*}

The choice of evaluation tasks is limited by the availability of labelled data, so most tasks involve transfer learning. Previously, we mentioned a simple linear regression task to predict the number of compact sources in a given crop, which helps determine training completion. This task is performed on MeerKAT data, using the available MGCLS and MIGHTEE compact source catalogs as labels. Therefore, the term "source count" simply refers to the total number of catalog-labelled sources in a single crop, as opposed to a stricter definition. In order to compare with previous works, we also evaluate binary classification of radio galaxy morphology using the public dataset MiraBest \citep{porter_mirabest_2023}. We extend this classification problem to hand-labelled FRI and FRII galaxies in the MIGHTEE Early Science fields, as well as multi-class classification.

Unless specified, these evaluation tasks are performed without any fine-tuning of the backbone; following our notation from Figure \ref{fig:framework}, the backbone $f_{\Phi}^*$ is fixed. Therefore, for speed and efficiency, we start by extracting the latent space embeddings $z$ for the dataset of interest $x$.
The final dense layer of a ResNet or the CLS token at the end of the transformer layers are typically used as a summary of the input image content. Because these embeddings, also commonly called features, are of different shapes depending on the network architecture, we use PCA to reduce the size to a vector of length 100\footnote{Interestingly, the fact that dimensionality reduction via PCA generally results in better performance indicates that features relevant to the chosen tasks occupy a small portion of the latent space.} 
before using them as input to the trainable regression or classification models $g_{\theta}$, which consist of a single linear layer. These regression and classification models were then trained for 100 and 300 epochs respectively, which thanks to the small shape of the input feature vectors took less than a minute on a single GPU.

This feature extraction approach, rather than attaching the classification or regression head directly, freezing the backbone, and performing the task, also allows us to easily compare with different backbones from the literature. The only modification necessary for extracting features was replication of the single-channel crops over the required number of channels (3 for backbones trained on natural images and 5 for optical Galaxy Zoo). 

All the backbones evaluated in this section are listed in Table \ref{tbl:backbones}; these naming conventions will be used from here onwards. The training weights of last two are not publicly available, so extracted features shared by \citet{mohale_enabling_2024} were used, and the metric reported for RGZ VDVAE originates directly from the paper of \citet{andrianomena_radio_2024}. All of these backbones were pre-trained \AN{in a self-supervised manner} on datasets at least an order of magnitude larger than \textit{MGCLS\_20k}, from DINOv2's 142 million images, to Galaxy Zoo and ImageNet's 1.2 million and Radio Galaxy Zoo's 108k images. 

All results in tabular form can be found in Appendix \ref{sec:results_table}.

\subsection{Compact source count prediction}\label{sec:linear}

Compact source count prediction is performed separately on both MGCLS and MIGHTEE crops. Feature vectors associated with source counts more than $3\sigma$ above or below the mean are considered outliers and not used for training or testing the model. This excludes 25 crops for MGCLS, and 4 for MIGHTEE. Selection of data by quality flag does not have a large effect on the chosen metric of mean squared error, so we do not exclude any crops for that reason. With the feature vectors extracted for each crop as input, the corresponding label is the total number of compact sources as per the catalog that are present in the crop. %This fact may indicate that image processing artefacts and source counts are independent of each other in latent space (as much as they can be since for pyBDSF source finding and noise are not independent).}

Despite the catalog-derived quantities being "noisy" labels, this simple regression task is a good test of the backbone's capabilities since it is a dataset-specific task, and can also be used as a linear probe to evaluate the network as it trains. 
%When evaluating during training or for the augmentation study, we report simply the MSE for a fit to the entire \textit{MGCLS\_20k} dataset. 

\begin{figure}
\includegraphics[width=.5\textwidth]{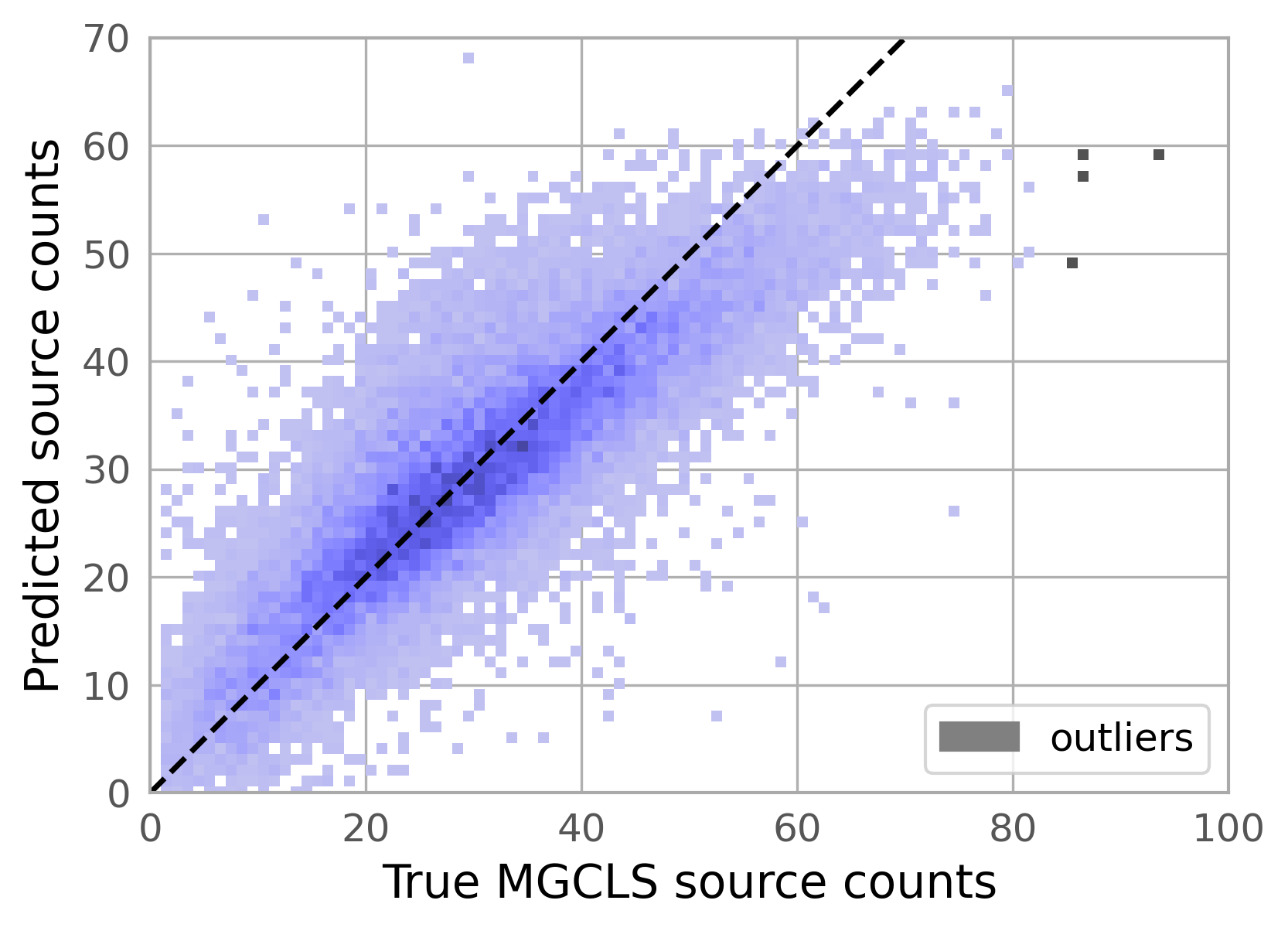}
\includegraphics[width=.5\textwidth]{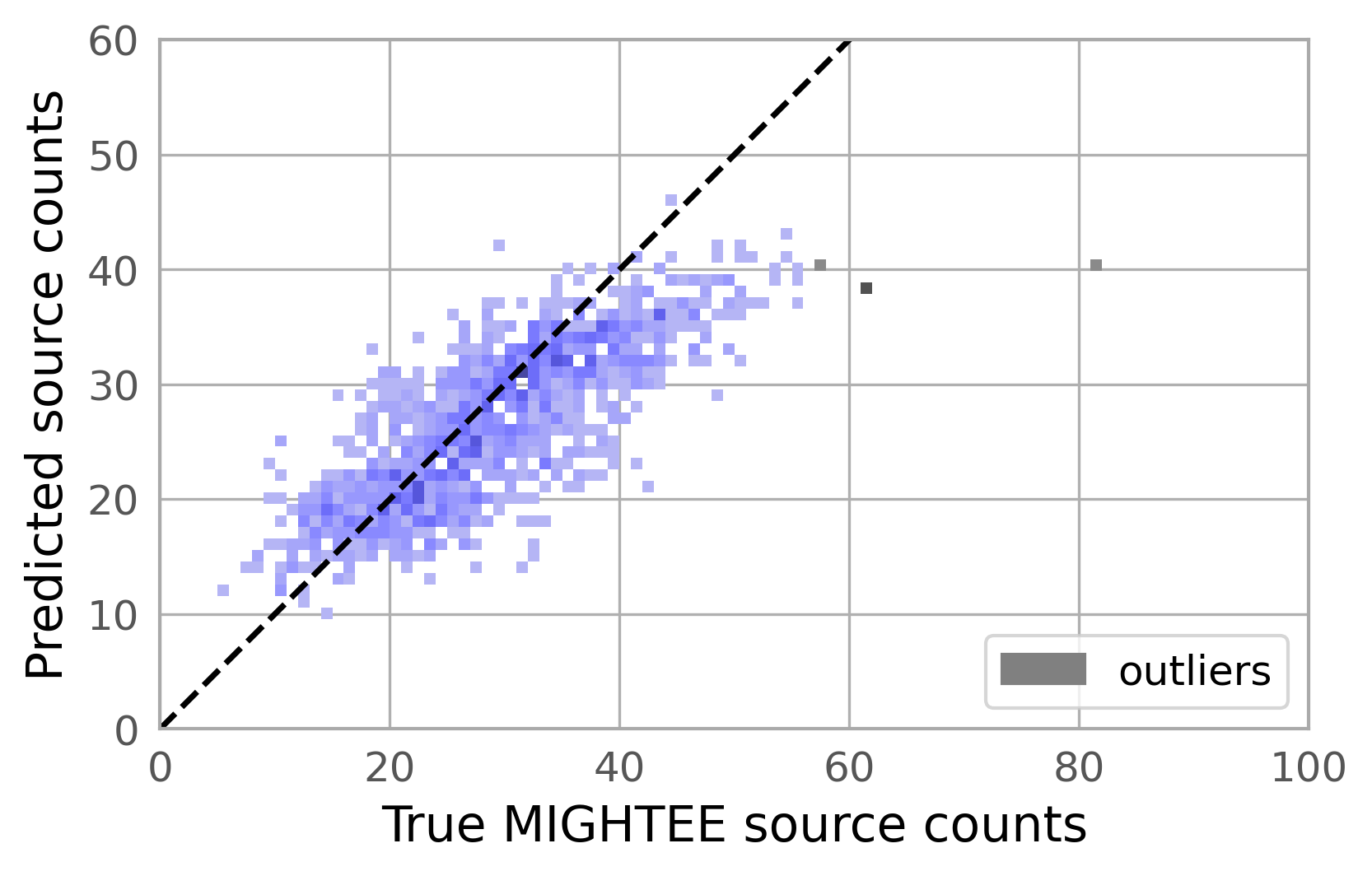}
\caption{Compact source count prediction via linear regression.}
\label{fig:mgclsregression}
\end{figure}

\AN{Similar to during training, we train and validate using a train/validation split of 70/30\% of the \textit{MGCLS\_20k} training dataset before evaluating on the \textit{MGCLS\_20k} test set, which has only crops not present in the \textit{MGCLS\_5k} subset.} Figure \ref{fig:mgclsregression} shows the results when evaluated on both the MGCLS and MIGHTEE datasets. %MSE for both MGCLS and MIGHTEE is reported in Table \ref{tbl:eval}. 
The reported errors are calculated on the results of three independent training runs. 
On MGCLS, the MGCLS Resnet backbone achieves a MSE of \AN{49.19 $\pm$ 0.06}, %after 425 epochs, 
while the MGCLS ViTS backbone does better with a MSE of \AN{39.98 $\pm$ 0.05}. %after 475 epochs. 
For MIGHTEE, the MSEs are 37.73 $\pm$ 0.01 and 23.14 $\pm$ 0.01, respectively. Due to its attention mechanism, ViT is generally better at learning properties of data, so this result is expected. 

\AN{Results in Figure \ref{fig:mgclsregression} indicate that the models tend to under-predict source counts in the upper few percentiles. For MGCLs, 2\% of crops have source counts exceeding 60, yet numbers greater than this were only predicted 0.15\% of the time. This behaviour is more evident with the MIGHTEE crops, where the upper $\sim$1\% of crops with more than 50 compact sources is always predicted to have much fewer, and no prediction on the MIGHTEE data exceeds 50 compact sources per crop. As the distribution of source counts in these datasets is not normal but rather skewed towards the mid-range and lower end, it is understandable that linear regression would struggle to properly fit the under-represented upper tail of the distribution.} 
%Nevertheless, perhaps} due to the higher sensitivity of the observation (2 $\mu$Jy vs 50 $\mu$Jy), MIGHTEE source counts were predicted with a lower MSE. \
The exact details of the source detection process might also have an impact, as already discussed in Section \ref{sec:labels}. 

\subsubsection{Regression: comparison with other backbones}\label{sec:counts_compare}

\begin{figure}
\includegraphics[width=\columnwidth]{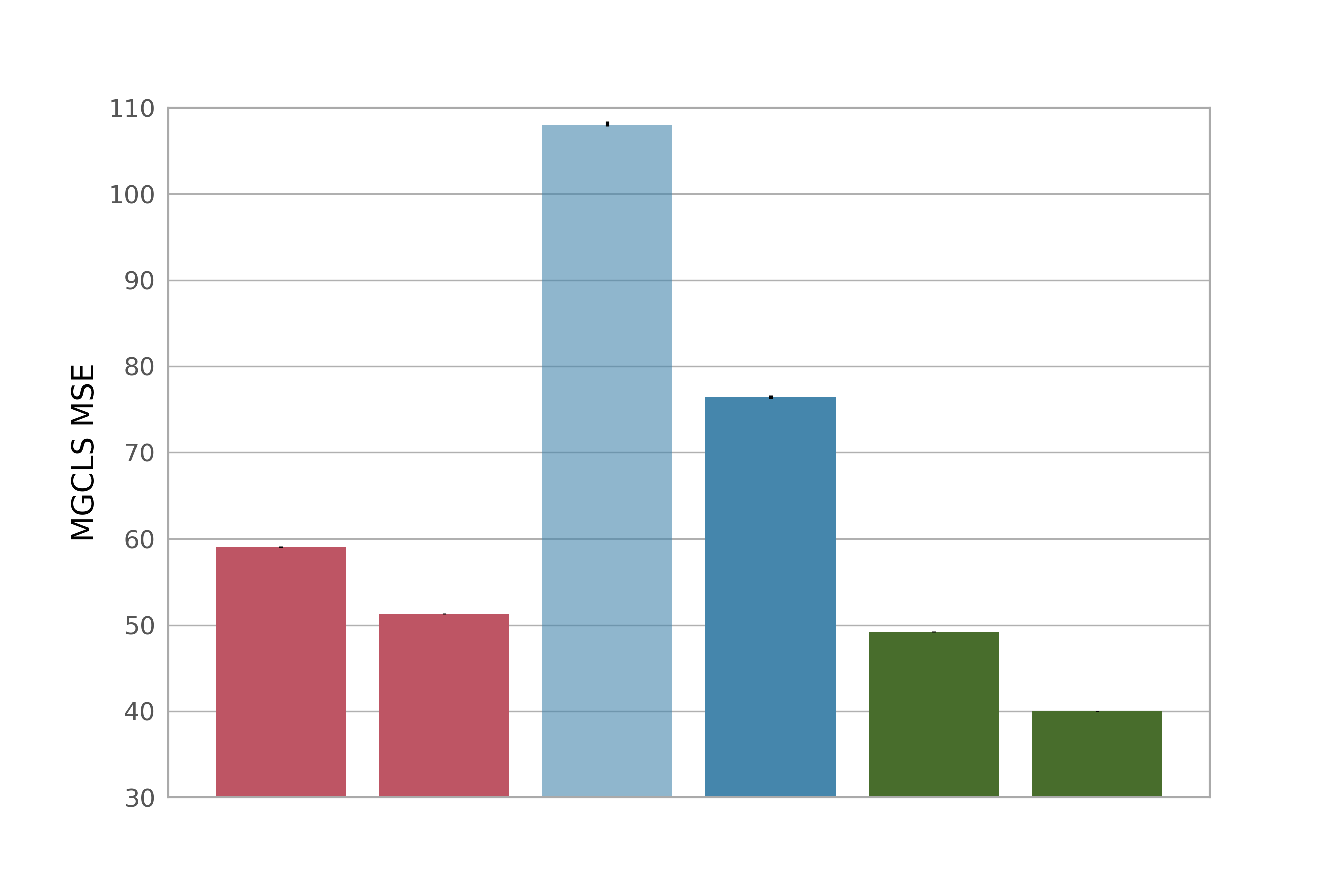}
\includegraphics[width=\columnwidth]{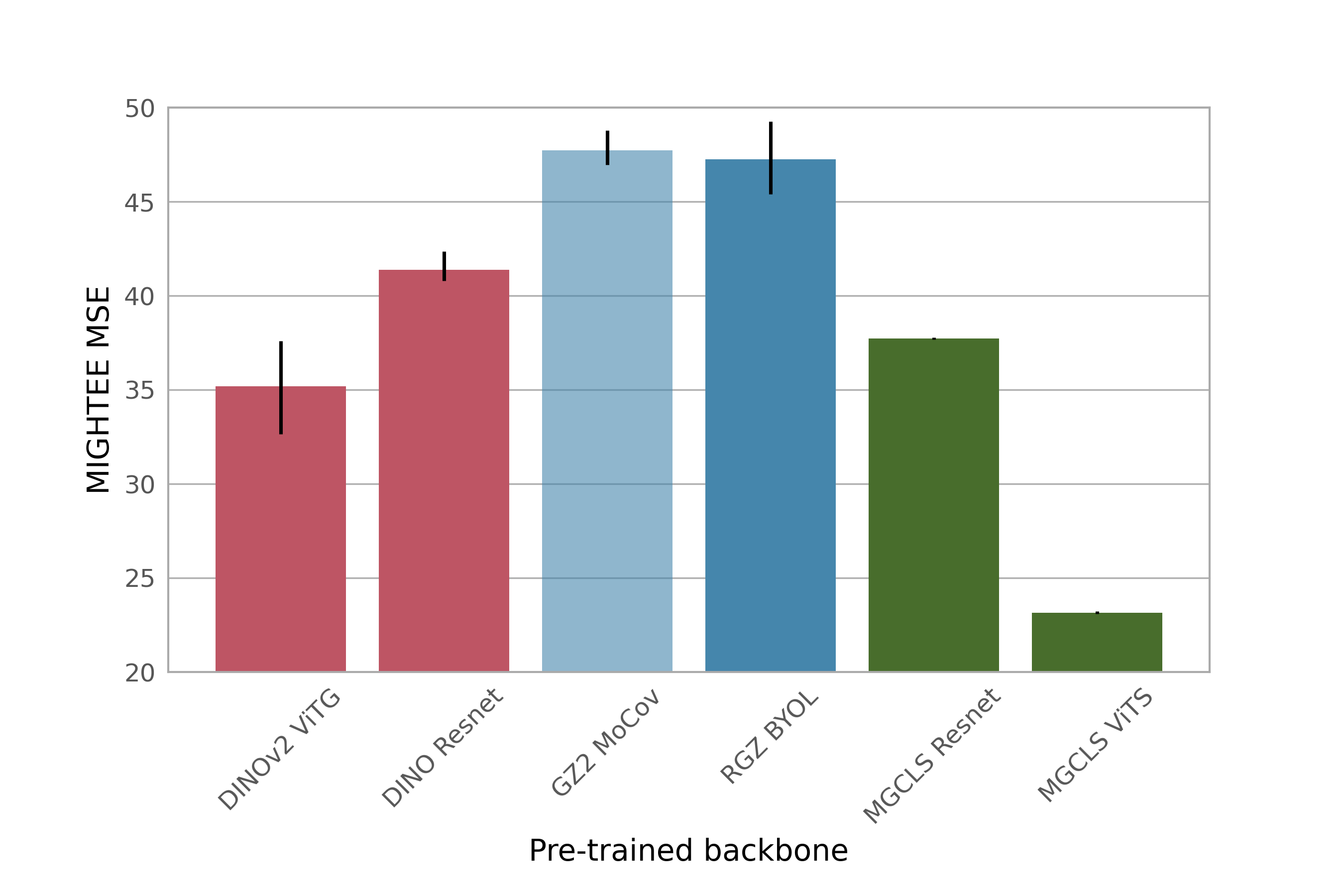}
\caption{Performance comparison of different backbones on the MGCLS (top) and MIGHTEE (bottom) compact source count linear regression task.}
\label{fig:regression}
\end{figure}

We compare our results with the current state-of-the-art in both computer vision and astrophysics. Figure \ref{fig:regression} illustrates how different backbones perform on compact source count prediction. It is clear that models trained with Galaxy Zoo datasets (GZ2 MoCo and RGZ BYOL) do not do well in this area, having not been exposed to images with such a variety of sources. It is interesting that the performance of DINOv2 VITG and DINO Resnet is much better than those models; perhaps the variety of images in their training data contributes to a more robust feature space. This is true when evaluated on the MIGHTEE observations as well; DINOv2's largest foundation model ViTG also transfers well to this particular dataset and task. Finally, our networks perform this task well, although that is of little surprise as they trained on the same data (with the exception MGCLS ViTS which did not see any of the evaluation data during training).

\subsection{FRI/FRII classification}\label{sec:classify}

Since the release of the MiraBest labelled dataset, it has become standard to evaluate networks on this binary classification task. In the Fanarhoff-Riley classification scheme, radio galaxies of type I are brighter towards the central galaxy or quasar, while type II are both generally brighter and brighter towards their lobes rather than the central component. Galaxy morphology can be difficult for humans to classify even with this straightforward paradigm, and there are many cases in which labelling is uncertain. We perform this classification using a single linear layer on both the MiraBest Confident dataset and \textit{MIGHTEE\_FR}. Both datasets are relatively class-balanced; MiraBest Confident has 397 FRI and 437 FRII, while \textit{MIGHTEE\_FR} has 90 FRI and 87 FRII galaxies.

In order to easily compare with \citet{slijepcevic_radio_2024}, %in Table \ref{tbl:eval} 
we report the test set error, which is one minus the test set accuracy. %, along with the F1 score and recall. 
MGCLS ResNet reaches a test set error of 10\% on MiraBest, while this is significantly worse for much smaller \textit{MIGHTEE\_FR} sample at 22\%. %\AN{elaborate on differences MiraBest/MIGHTEE} 
The MGCLS ViTS backbone performs slightly worse, with a MiraBest test set error of 13.5\% and 32\% for \textit{MIGHTEE\_FR}. Using our evaluation pipeline of feature extraction, PCA, and classification with a single linear layer, the test set error for RGZ BYOL is 11\% and 42\% for \textit{MIGHTEE\_FR}. This demonstrates that our model, trained with 20K source-rich crops from wide-field images, can learn the same significant features as a model trained with 108K images from Radio Galaxy Zoo. %for this use case anyway

\subsubsection{Classification: comparison with other backbones}\label{sec:fr_compare}

\begin{figure}
\includegraphics[width=.5\textwidth]{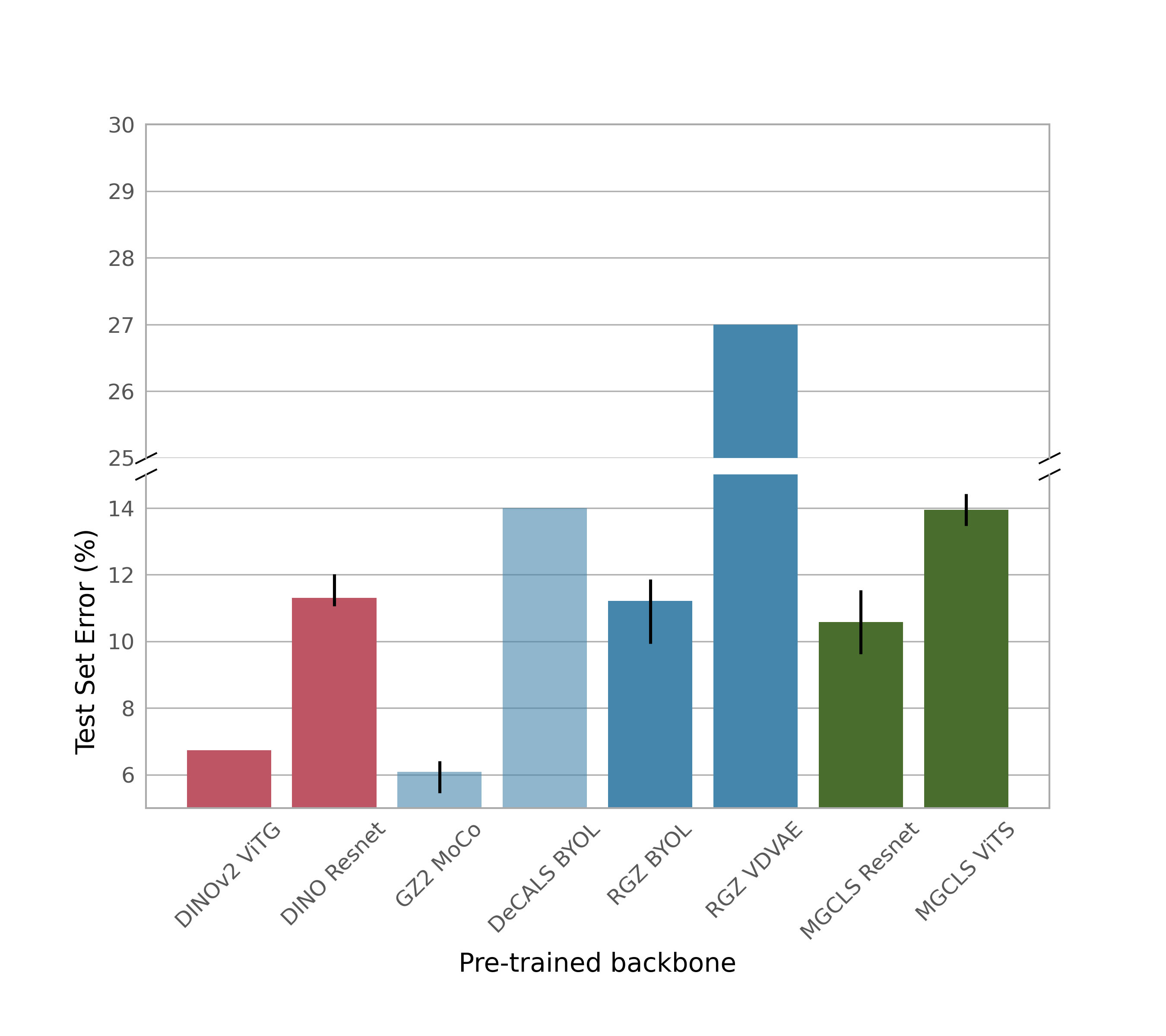}
\caption{Performance comparison of different backbones with FRI/FRII classification on MiraBest Confident.}
\label{fig:classification}
\end{figure}

Figure \ref{fig:classification} shows how binary classification of radio galaxies is easily done by models pre-trained on large datasets. Both DINOv2 ViTG and RGZ MoCo out-perform the state-of-the-art in the radio astronomy literature; ImageNet-trained DINO Resnet comes very close to the RGZ BYOL result. %Architecture and training details can also be seen to have an effect, but in general both amount and variety of training data ; 

If a model is considered versatile because it is able to adapt well to more than one task, then both DINOv2 ViTG and our models can be considered closer to foundation models than those trained on Galaxy Zoo. These results contradict the \AN{frequently cited} need for domain-specific training data, since a 1,100-million parameter model trained on 142 million images out-performs the state-of-the-art model trained on Radio Galaxy Zoo.

However, we must acknowledge that neither source count prediction nor binary FRI/FRII classification are complex tasks. Standard benchmark tasks in computer vision literature, such as 1000-class classification with ImageNet, demand more flexibility from both backbone and classifier. A third class of hybrid galaxies is available in MiraBest, but there are only 19 confidently-labelled samples, so we evaluate multi-class classification on a different dataset from the same survey.

 \subsection{Multiclass classification with fine-tuning}\label{sec:multiclass}

\begin{figure}
\includegraphics[width=.5\textwidth]{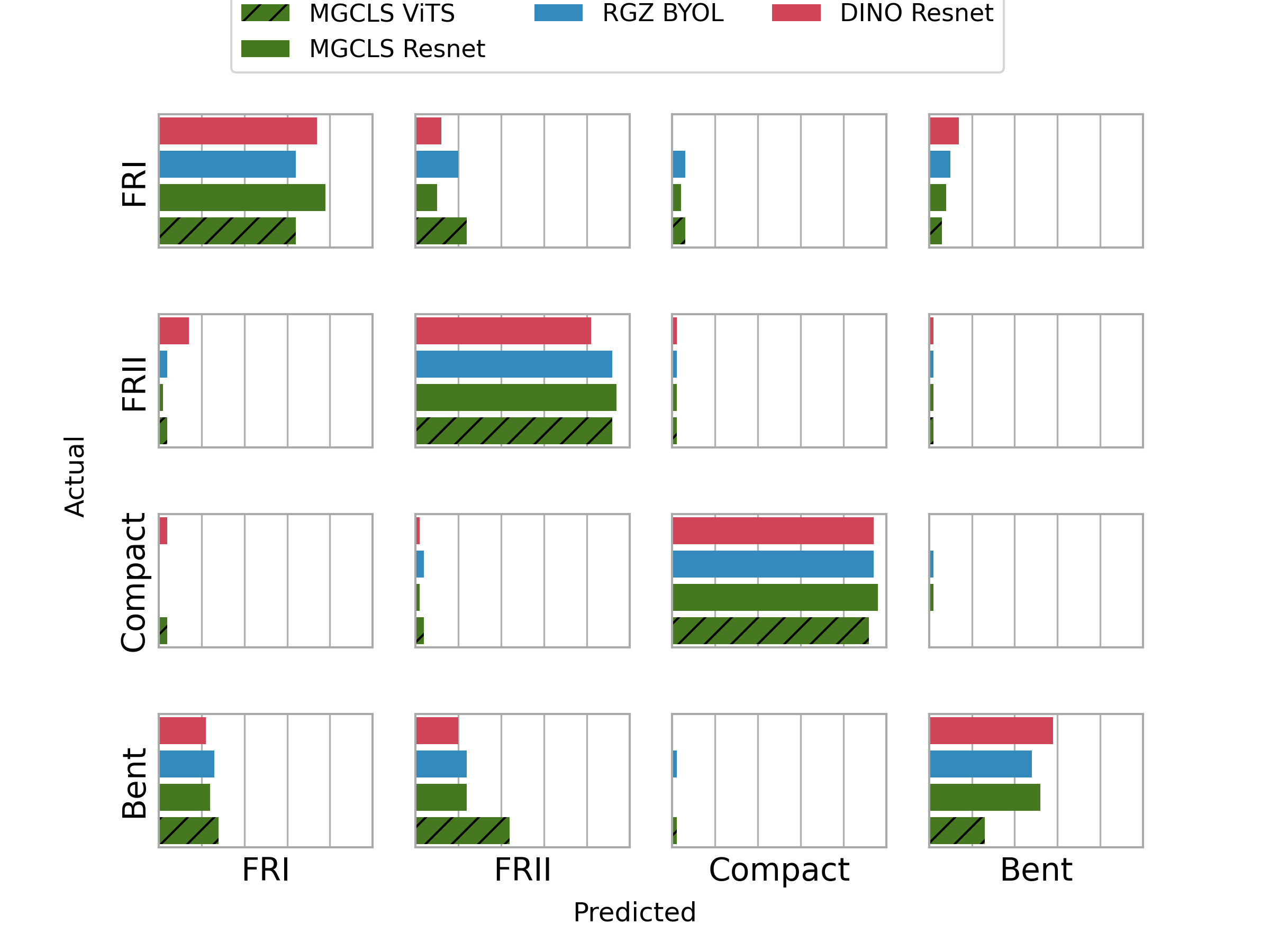}
\caption{Confusion matrix comparison of different backbones with multiclass morphology classification on RadioGalaxyDataset \citep{griese_floriangrieseradiogalaxydataset_2022}. Each class in the test set has 50 samples.}
\label{fig:multiclass}
\end{figure}

 In order to demonstrate adaptability to more complex tasks, we make use of \citet{griese_floriangrieseradiogalaxydataset_2022}'s RadioGalaxyDataset, a collection of FIRST survey images of four morphology types. Besides FRI and FRII, compact sources and bent-tailed galaxies are included. In total, there are 495 FRI, 924 FRII, 391 compact, and 348 bent galaxies. In the validation and test sets, the classes are perfectly balanced with 50 samples of each class.

 Classification results using a single layer are poor for every class except compact sources, so fine-tuning is a necessity for this more complex task. In addition to comparing our models with RGZ BYOL, we also show the results of evaluating DINO Resnet. All layers of the ResNet backbones are allowed to fine-tune, as is the entire Vision Transformer backbone. %Results for all fine-tuneable networks are tabulated in Appendix \ref{sec:results_table}. maybe just have a footnote saying GZ2 was bad? or even 
 
 The best results out of three evaluation runs for each backbone are presented in Figure \ref{fig:multiclass}. All networks are capable of accurately identifying compact sources, although RGZ BYOL and MGCLS ViTS tend to misclassify these more often than the others. The addition of the bent-tailed galaxy class causes some confusion between that and the FRI/FRII morphologies, which can also feature extended emission. This is evident in the way that DINO Resnet classifies bent-tailed galaxies better than the others, but at the cost of FRI/FRII classification accuracy. Similarly, MGCLS ViTS errs in favor of FRII galaxies, at the cost of distinguishing bent and FRI galaxies poorly.

 The end result, however, is only minor differences between the performances of different pre-trained ResNet backbones, after fine-tuning. %but ViT is worse...

\section{Summary}\label{sec:summary}

In this work, we demonstrate that source-rich crops from wide-field MeerKAT continuum images, which require minimal effort in selection and pre-processing, can be used to train a multipurpose backbone. The DINO framework for self-supervised learning, which encourages the backbone network to learn local-to-global correspondences within the data, lends itself especially well to these images as opposed to sparser, galaxy-centered ones. The results we achieve, depending on the evaluation task, are similar to or better than the current state-of-the-art. 

This result has profound impacts on scalability of domain-specific foundation models. Data straight from the calibration and imaging pipelines of modern telescopes can immediately be incorporated into such models, eliminating the need to wait for millions of single-galaxy cutouts to be collected into a Galaxy Zoo before applying self-supervised learning. Furthermore, a relatively small number of training samples may be sufficient. Our training dataset of twenty thousand crops is an order of magnitude smaller than Radio Galaxy Zoo, and even training with a five thousand sample subset of that data results in good performance on our chosen evaluation tasks. Many radio telescopes already have archived collections of pipeline data products that are at least of this magnitude; if our findings generalize well across different bandwidths and to observatory-specific tasks, then instrument-specific fine-tuned models are easily within reach.

%do we need domain-specific training? or are natural-image trained models good enough
Our findings also show that publicly-available foundation models trained on natural images can perform radio astronomy classification and regression tasks, sometimes even better than models trained with astrophysical images. It is true that domain-specific data reduces by orders of magnitude the amount of training data and computing resources needed to achieve similar performance to the largest computer vision foundation models. From a practical perspective however, it is extremely easy to initiate a network with the weights from pre-trained open-source models, so these remain a good starting point for approaching astrophysics tasks with deep learning. Even when allowing networks to fine-tune, the difference between natural image backbones and domain-specific ones is on the order of a few percent (see for example Table \ref{tbl:layers}). 

However, these small advantages might become more significant once tasks become more complex and specific to astronomy or radio astronomy. The standard task of binary morphology classification is relatively simple, and even the additional source count task we use is easily done by linear regression. Other examples in the literature evaluate similarity search, although almost always qualitatively. The small variety of downstream tasks is due in part to a lack of labelled data. \citet{hayat_self-supervised_2021} were able to demonstrate performance on a number of different binary classification tasks as well as redshift estimation, thanks to the labelled Galaxy Zoo dataset. For tasks independent of channel, backbones trained with radio astronomy data could be evaluated on optical astronomy datasets; it has already been done in reverse, e.g. in \citet{mohale_enabling_2024}. With the emergence of new labelled datasets such as \citet{gupta_radiogalaxynet_2024}, tasks like source detection and parameter estimation should also become standard. It will be interesting to see if a complex, domain-specific task like this will clearly show the need for domain-specific training data or not.

In addition to adapting easily and accurately to perform tasks that are currently done by analytic pipelines, foundation models in radio astronomy should ideally be sensitive to features that allow room for discovery. This includes diffuse emission, which tends to be more evident in lower-resolution image reconstructions than the high-resolution ones which favor compact sources.
The MGCLS survey alone contains many examples of diffuse emission, from radio relics spanning megaparsecs to mini-halos only ten to hundreds of kiloparsecs in size. Even the small sky area observed by the MIGHTEE survey has giant radio galaxies \citep{delhaize_mightee_2021}, one of which required a customized CLEAN weighting to properly image diffuse emission from the galaxy lobes. Models which are able to take full advantage of the high-dynamic range, wide-field images of telescopes like MeerKAT and ASKAP will almost certainly be more capable of discovery than standard computer vision models. 

%and to get away from images and to UV
There are still many ways to leverage the uniqueness of radio astronomy data that have yet to be thoroughly explored.
In the image plane, the inclusion of physics-driven data augmentations may turn out to be key; %, as it has in \AN{other fields, find examples/citations}. 
or the dependence on optimal augmentation schemes could simply be bypassed by using masking. 
With radio astronomy, there are also other modalities of the data to explore, such as spectral and temporal, or different stages of data processing at which self-supervised learning could be applied. Images directly from the automated processing pipelines, such as the MGCLS "basic" CLEAN images, dirty images more closely resembling the Fourier inversion of the measured visibilities, or the calibrated visibilities themselves are all representations of the same observation. Abandoning the reconstructed images in favor of the Fourier plane visibilities brings its own challenges, especially with data preparation. Some recent successes with source localization \citep{taran_challenging_2023} and reconstruction \citep{vafaeisadr_deepsource_2019, drozdova_radio-astronomical_2023} show promise in this area, an exciting direction for future research. 

\section*{\AN{Data Availability Statement}}\label{sec:data_availability}
\AN{All datasets used in this work are public, and the code and pre-trained checkpoints are available at \url{https://github.com/elastufka/mgcls_dino}.}

 \section*{Acknowledgements}\label{sec:acknowledgements}
 
This work has been done in partnership of the
Swiss SKA consortium which is funded by the State Secretariat for Education, Research and Innovation (SERI).

MGCLS data products were provided by the South African Radio Astronomy Observatory and the MGCLS team and were derived from observations with the MeerKAT radio telescope. The MeerKAT telescope is operated by the South African Radio Astronomy Observatory, which is a facility of the National Research Foundation, an agency of the Department of Science and Innovation.

The authors would like to thank Prof. Anna Scaife and team for sharing their FRI/FRII classifications on the MIGHTEE survey, and Koketso Mohale for sharing extracted features from the work in \cite{mohale_enabling_2024}.

%OB is supported by the {\em AstroSignals} Sinergia Project funded by the Swiss National Science Foundation.

OB and DP were supported by the SNF Sinergia grant CRSII5-193826 ``AstroSignals: A New Window on the Universe, with the New Generation of Large Radio-Astronomy Facilities''.

MD and VK were supported by the SNF Sinergia project (CRSII5-193716), ``Robust deep density models for high-energy particle physics and solar flare analysis (RODEM)''.

\bibliographystyle{aa}
\bibliography{meerkat_ssl_paper}

\begin{appendix} %these should go after the bibliography
  \onecolumn
\section{MGCLS Data Quality illustration}\label{sec:dqf}

\begin{figure*}
\includegraphics[width=0.88\paperwidth]{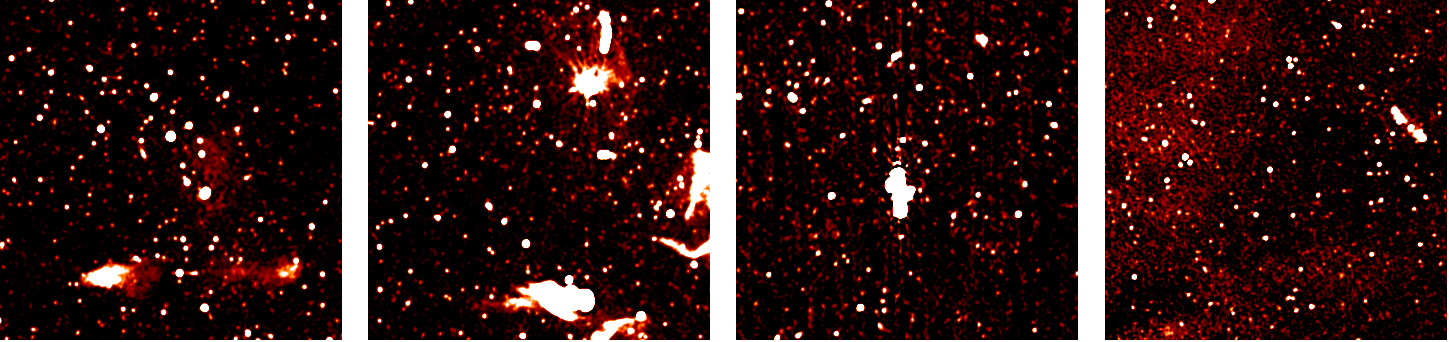}
\caption{\AN{600$\times$600 pixel crops from various MGCLS fields illustrating the Data Quality Flags (DQF). From left to right, they are: 0 = good dynamic range; 1 = moderate dynamic range with some artefacts around bright sources; 2 = poor dynamic range with high contamination by bright source artefacts; 3 = poor dynamic range with ripples across image.}} %Backbones pre-trained on radio astronomy data out-perform the backbone initialized with ImageNet weights.}
\label{fig:dqf}
\end{figure*}

\AN{Figure \ref{fig:dqf} illustrates MGCLS images of various data quality. From left to right, they are: Abell 22, with DFQ 0, Abell 85 with DQF 1, Abell 168 wiht DQF 2, and J1653.0-5943 with DQF 3. Images are displayed without scaling, within the flux range of 1e-7 to 5e-5 mJy/beam.}

\section{Evaluation results}\label{sec:results_table}

\begin{table*}[h]
\caption{Results of evaluation tasks from Section \ref{sec:evaluation}}\label{tbl:results}
\begin{tabular}{p{0.12\paperwidth} p{0.11\paperwidth} p{0.11\paperwidth} p{0.05\paperwidth}p{0.05\paperwidth}p{0.05\paperwidth}p{0.1\paperwidth}p{0.1\paperwidth}}
%{@{}p{0.12\pagewidth} p{0.2\pagewidth} p{0.15\pagewidth} p{0.08\pagewidth}p{0.08\pagewidth}p{0.1\pagewidth}}%[width=\pagewidth]
\textbf{Backbone Name} & \textbf{MGCLS MSE} & \textbf{MIGHTEE MSE} & \textbf{MB error} & \textbf{MB F1} & \textbf{MFR error} & \textbf{MFR F1} & \textbf{RGD \nobreak{}Accuracy}\\
\hline  & \\[-1.5ex]
% MGCLS Resnet   & 97.133 $\pm$	0.0005  &  23.145000 $\pm$	0.068000  & 0.105800 $\pm$	0.009600   & 0.8989 0.01389 & 0.219048 	0.014286 & & \\
% MGCLS ViTS   & 93.537 $\pm$	0.005   &   	37.731000 $\pm$	0.059500 & 0.139400 $\pm$	0.004800    & 0.8475 0.01 & 0.324800 	0.012800 & 0.7111 \\
% RGZ BYOL  &  155.751592 $\pm$	0.038185  & 47.236032 $\pm$	1.935560 & 0.112179 $\pm$	0.009615     &  0.8923 0.01     & 0.285714 	0.001350 &  \\
% GZ2 MoCo  &  161.038 $\pm$	0.551765  &  47.712498 $\pm$	0.908606 & 0.060897 $\pm$	0.004808   & 0.9388 0.0059  &  	0.375510 	0.042857 &   \\
% DINO Resnet  &  104.545 $\pm$	0.182  &  41.385000 $\pm$	0.787500 & 0.113000 $\pm$	0.004800   & 0.8962 0.0043     & 0.314286 	0.000000 & \\
% DINOv2 ViTG  &  113.081965 $\pm$	0.004967  &  35.180940 $\pm$	2.474007 & 0.067308 $\pm$	0.000000  & 0.9363     & 0.300000 	0.014286 & \\
% DeCALS BYOL  &  -  &  -   & 0.148400 	0.004050     & 0.8513 0.0013 \\
% RGZ VDVAE  &  - &  - & 0.270000  & -     & - \\
%Astronomaly   & MGCLS enhanced images     &  Anomaly detection     & 9582 \\                
DINOv2 ViTG & 59.08 ± 0.07 & 32.79 ± 0.00 & 0.07 & 0.94 & 0.33 & 0.72 ± 0.00 & - \\
DINO Resnet & 51.34 ± 0.06 & 41.39 ± 0.19 & 0.11 & 0.90 & 0.27 & \textbf{0.75 ± 0.00} & 0.76 ± 0.01 \\
GZ2 MoCo & 107.98 ± 0.29 & 47.10 ± 0.07 & \textbf{0.06} & \textbf{0.94} & 0.54 & 0.48 ± 0.00 & 0.70 ± 0.01 \\
RGZ BYOL & 76.39 ± 0.21 & 47.24 ± 0.08 & 0.11 & 0.89 & 0.42 & 0.59 ± 0.00 & 0.79 ± 0.02 \\
MGCLS Resnet & 49.19 ± 0.06 & 37.73 ± 0.01 & 0.10 & 0.90 & \textbf{0.22} & 0.75 ± 0.01 & \textbf{0.80 ± 0.00} \\
MGCLS ViTS & \textbf{39.98 ± 0.05} & \textbf{23.14 ± 0.01} & 0.15 & 0.86 & 0.32 & 0.74 ± 0.00 & 0.68 ± 0.01 \\
DeCALS BYOL  &  -  &  -   & 0.15   & 0.85  & - & -\\ %.15 $\pm$	0.004050 & 0.8513 0.0013 
RGZ VDVAE  &  - &  - & 0.27  & -     & - & - & - \\ 
\hline & \\[-1.5ex]
\end{tabular}\tablefoot{Datasets are abbreviated as follows: \textbf{M}ira\textbf{B}est, \textbf{M}IGHTEE\_\textbf{FR}, and \textbf{R}adio\textbf{G}alaxy\textbf{D}ataset. The error quantity is test set error. Uncertainties are less than 1\% if not included. }
\end{table*}

\section{Fine-tuning and label reduction results}

\subsection{Fine-tuning}\label{sec:mbfinetune}

For a more in-depth comparison with \citet{slijepcevic_radio_2024} and to show how performance on binary classification can improve, we perform fine-tuning of all available Resnet backbones. The classification head attached to the backbone consists of a single input layer of shape (\textit{embedding\_dimension}, 512) and an output layer of shape (512, \textit{n\_classes}). This is to account for the different shape of the embedding layers between ResNets of different depths, such as our Resnet50 and RGZ BYOL's Resnet18. Additionally, we use a slightly larger center crop in both the training and validation transforms (96 vs 70 pixels) and do not include an additional random small crop or Gaussian blur in the training transformations. Therefore, even though our training hyperparameters are similar, we do not reproduce the exact results reported in \cite{slijepcevic_radio_2024}.

Results showing test set error as a function of the number of fine-tuning layers is shown in Figure \ref{fig:labels}a. For clarity, we only plot the fine-tuning results of our work, MGCLS Resnet, and RGZ BYOL; complete results are reported in Table \ref{tbl:layers}. %For context, the test set error of a Resnet50 backbone loaded with DINO's default weights trained from ImageNet is also plotted. 
Error bars are determined by the accuracies reported during the last ten epochs of training, rather than from the results of multiple training runs, due to the less efficient nature of fine-tuning.  

%The difference between the performance of various backbones is at most 4\%; however, both backbones pre-trained with radio astronomy data do outperform the DINO default. 
Unlike the findings of \citet{slijepcevic_radio_2024}, we do not find significant benefit from fine-tuning all model layers. The difference in training data augmentation schemes may be key in this case; focusing even more strongly on the bright galaxy at the center of the image could result in better classification. 

\subsection{Reducing number of labels}\label{sec:nlabels}

\begin{figure}[h]
  \includegraphics[width=.43\paperwidth]{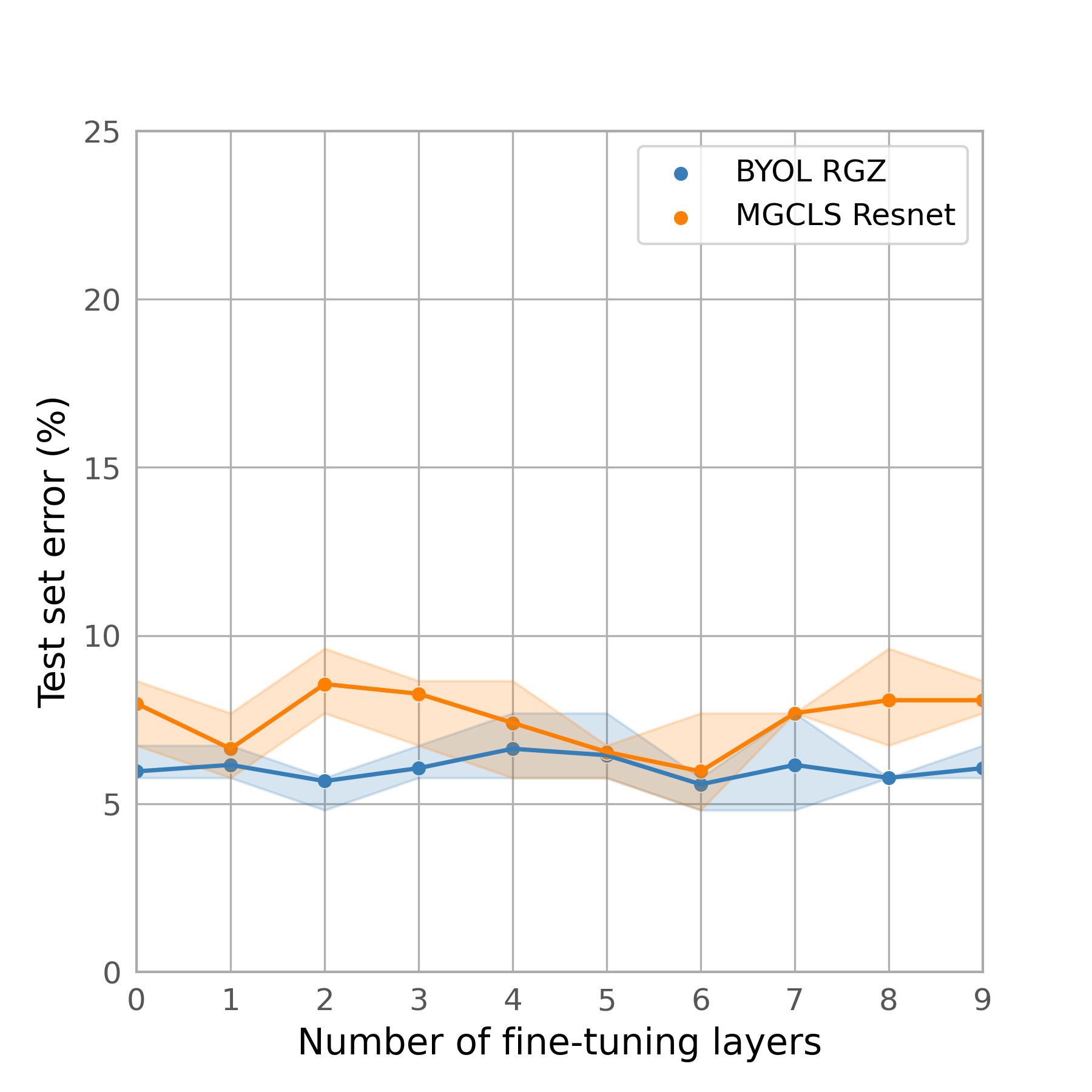}
  \includegraphics[width=.43\paperwidth]{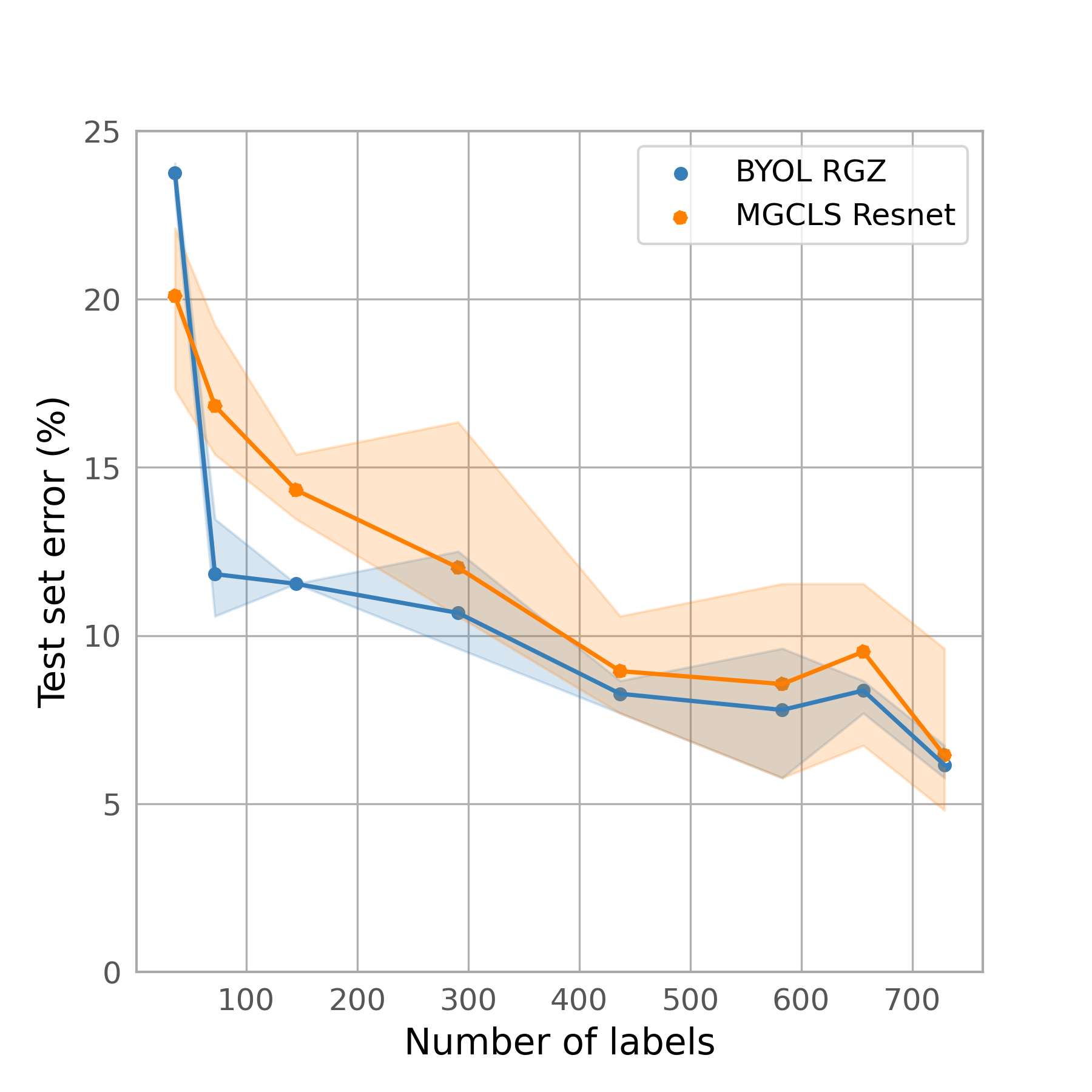}
  \caption{Left: MiraBest Confident test set error according to number of labels present in the training data. As expected, performance improves significantly with more labeled data. Right: MiraBest Confident test set error according to number of fine-tuning layers. With our chosen training data augmentations and hyperparameters, performance does not improve significantly with additional depth in the model.}
  \label{fig:labels}
  \end{figure}

The impact that the number of labeled samples used for training can have is already evident from our single-layer FRI/FRII classification with \textit{MIGHTEE\_FR}, resulting in test set error of $\sim$20 - 30\%. With only 140 labeled samples in the training set, we can expect significant improvement with more data. 

This is certainly the case with MiraBest; by steadily increasing the number of labels used for training, the accuracy of the classifier also increases. Figure \ref{fig:labels}b shows the same trend as the result shown in \citet{slijepcevic_radio_2024}'s figure 2. When even half the available labels are used, the accuracy of MGCLS Resnet backbone quickly approaches that RGZ BYOL. Complete results are reported in Table \ref{tbl:labels}.

\twocolumn
\begin{table}
\begin{tabular}{p{0.12\paperwidth} p{0.06\paperwidth} p{0.08\paperwidth} p{0.08\paperwidth}}
\textbf{Backbone Name} & \textbf{Fine-tuned Layers} & \textbf{MB error} & \textbf{MB F1} \\
\hline  & \\[-1.5ex] 
\multirow[c]{10}{*}{DINO Resnet} & 0 & 0.09 ± 0.01 & 0.92 ± 0.01 \\
 & 1 & 0.07 ± 0.01 & 0.93 ± 0.01 \\
 & 2 & 0.08 ± 0.01 & 0.92 ± 0.01 \\
 & 3 & 0.09 ± 0.01 & 0.92 ± 0.01 \\
 & 4 & 0.07 ± 0.00 & 0.93 ± 0.00 \\
 & 5 & 0.08 ± 0.01 & 0.92 ± 0.01 \\
 & 6 & 0.09 ± 0.00 & 0.91 ± 0.01 \\
 & 7 & 0.08 ± 0.00 & 0.93 ± 0.00 \\
 & 8 & 0.07 ± 0.00 & 0.93 ± 0.00 \\
 & 9 & 0.09 ± 0.01 & 0.92 ± 0.01 \\
  \hline  & \\[-1.5ex] 
\multirow[c]{10}{*}{GZ2 MoCo} & 0 & 0.09 ± 0.00 & 0.91 ± 0.00 \\
 & 1 & 0.08 ± 0.00 & 0.93 ± 0.00 \\
 & 2 & 0.09 ± 0.01 & 0.92 ± 0.01 \\
 & 3 & 0.10 ± 0.00 & 0.91 ± 0.00 \\
 & 4 & 0.09 ± 0.01 & 0.92 ± 0.00 \\
 & 5 & 0.09 ± 0.01 & 0.92 ± 0.00 \\
 & 6 & 0.09 ± 0.00 & 0.92 ± 0.00 \\
 & 7 & 0.08 ± 0.01 & 0.93 ± 0.01 \\
 & 8 & 0.08 ± 0.00 & 0.93 ± 0.00 \\
 & 9 & 0.08 ± 0.01 & 0.93 ± 0.01 \\
  \hline  & \\[-1.5ex] 
\multirow[c]{10}{*}{RGZ BYOL} & 0 & 0.06 ± 0.00 & 0.95 ± 0.00 \\
 & 1 & 0.06 ± 0.01 & 0.94 ± 0.00 \\
 & 2 & 0.06 ± 0.00 & 0.95 ± 0.00 \\
 & 3 & 0.06 ± 0.00 & 0.94 ± 0.00 \\
 & 4 & 0.07 ± 0.01 & 0.94 ± 0.01 \\
 & 5 & 0.07 ± 0.01 & 0.94 ± 0.01 \\
 & 6 & 0.06 ± 0.00 & 0.95 ± 0.00 \\
 & 7 & 0.06 ± 0.01 & 0.94 ± 0.01 \\
 & 8 & 0.06 ± 0.00 & 0.95 ± 0.00 \\
 & 9 & 0.06 ± 0.00 & 0.94 ± 0.00 \\
\hline  & \\[-1.5ex] 
\multirow[c]{10}{*}{MGCLS Resnet} & 0 & 0.08 ± 0.01 & 0.93 ± 0.01 \\
 & 1 & 0.07 ± 0.01 & 0.94 ± 0.01 \\
 & 2 & 0.09 ± 0.01 & 0.92 ± 0.01 \\
 & 3 & 0.09 ± 0.00 & 0.92 ± 0.00 \\
 & 4 & 0.08 ± 0.01 & 0.93 ± 0.01 \\
 & 5 & 0.07 ± 0.00 & 0.94 ± 0.00 \\
 & 6 & 0.06 ± 0.01 & 0.94 ± 0.01 \\
 & 7 & 0.08 ± 0.00 & 0.93 ± 0.00 \\
 & 8 & 0.08 ± 0.01 & 0.92 ± 0.01 \\
 & 9 & 0.07 ± 0.00 & 0.94 ± 0.00 \\
 \hline  & \\[-1.5ex] 
\end{tabular}\caption{Performance of each backbone on binary morphology classification with MiraBest Confident, as the number of fine-tuned layers in the ResNet increases. }\label{tbl:layers}
\end{table}

\begin{table}
\begin{tabular}{p{0.12\paperwidth} p{0.06\paperwidth} p{0.08\paperwidth} p{0.08\paperwidth}}
\textbf{Backbone Name} & \textbf{Number of \nobreak{}Labels} & \textbf{MB error} & \textbf{MB F1} \\
\hline  & \\[-1.5ex] 
\multirow[c]{8}{*}{DINO Resnet} & 36 & 0.29 ± 0.01 & 0.76 ± 0.01 \\
 & 72 & 0.15 ± 0.00 & 0.87 ± 0.00 \\
 & 145 & 0.11 ± 0.01 & 0.90 ± 0.01 \\
 & 291 & 0.11 ± 0.01 & 0.90 ± 0.01 \\
 & 437 & 0.08 ± 0.01 & 0.93 ± 0.01 \\
 & 583 & 0.05 ± 0.01 & 0.95 ± 0.01 \\
 & 656 & 0.07 ± 0.01 & 0.93 ± 0.01 \\
 & 729 & 0.09 ± 0.01 & 0.92 ± 0.01 \\
  \hline  & \\[-1.5ex]
 \multirow[c]{8}{*}{GZ2 MoCo} & 36 & 0.29 ± 0.00 & 0.77 ± 0.00 \\
 & 72 & 0.17 ± 0.01 & 0.85 ± 0.00 \\
 & 145 & 0.15 ± 0.01 & 0.86 ± 0.01 \\
 & 291 & 0.13 ± 0.02 & 0.88 ± 0.01 \\
 & 437 & 0.12 ± 0.01 & 0.89 ± 0.01 \\
 & 583 & 0.09 ± 0.01 & 0.92 ± 0.01 \\
 & 656 & 0.09 ± 0.01 & 0.92 ± 0.01 \\
 & 729 & 0.08 ± 0.00 & 0.93 ± 0.00 \\
  \hline  & \\[-1.5ex]
\multirow[c]{8}{*}{BYOL RGZ} & 36 & 0.22 ± 0.01 & 0.82 ± 0.01 \\
 & 72 & 0.14 ± 0.01 & 0.88 ± 0.01 \\
 & 145 & 0.12 ± 0.01 & 0.89 ± 0.01 \\
 & 291 & 0.12 ± 0.01 & 0.89 ± 0.01 \\
 & 437 & 0.09 ± 0.01 & 0.91 ± 0.01 \\
 & 583 & 0.08 ± 0.01 & 0.93 ± 0.01 \\
 & 656 & 0.09 ± 0.01 & 0.91 ± 0.01 \\
 & 729 & 0.07 ± 0.01 & 0.93 ± 0.01 \\
 \hline  & \\[-1.5ex]
\multirow[c]{8}{*}{MGCLS Resnet} & 36 & 0.20 ± 0.02 & 0.81 ± 0.02 \\
 & 72 & 0.18 ± 0.01 & 0.84 ± 0.01 \\
 & 145 & 0.14 ± 0.01 & 0.86 ± 0.01 \\
 & 291 & 0.11 ± 0.02 & 0.90 ± 0.02 \\
 & 437 & 0.09 ± 0.01 & 0.91 ± 0.01 \\
 & 583 & 0.09 ± 0.02 & 0.92 ± 0.02 \\
 & 656 & 0.11 ± 0.01 & 0.90 ± 0.01 \\
 & 729 & 0.06 ± 0.01 & 0.94 ± 0.01 \\
 \hline  & \\[-1.5ex] 
\end{tabular}\caption{Performance of each backbone on binary morphology classification with MiraBest Confident, as the number of labeled samples in the training set increases.}\label{tbl:labels}
\end{table}

\end{appendix}

\end{document}